\documentclass[10pt,journal,compsoc]{IEEEtran}
\usepackage{colortbl}
\usepackage[utf8]{inputenc} % allow utf-8 input
\usepackage[T1]{fontenc}    % use 8-bit T1 fonts
\usepackage{url}            % simple URL typesetting
\usepackage{booktabs}       % professional-quality tables
\usepackage{amsfonts}       % blackboard math symbols
\usepackage{nicefrac}       % compact symbols for 1/2, etc.
\usepackage{microtype}      % microtypography
\usepackage[dvipsnames, svgnames, x11names,table]{xcolor}

\definecolor{darkamber}{RGB}{195, 95, 0}
\usepackage[pagebackref=true,breaklinks=true,colorlinks,citecolor=ForestGreen]{hyperref}
\usepackage{graphicx}
\usepackage{tabularx}
\usepackage{amsmath}
\usepackage{subfigure}
\usepackage{tcolorbox}
\usepackage{multirow}
\usepackage{pifont}
\usepackage{enumitem}
\usepackage{amssymb}

\usepackage{enumitem}
\usepackage[misc]{ifsym}
\usepackage{bm}
\usepackage{multirow}
\usepackage{bbding}
\usepackage{array}
\usepackage{hyperref}
\definecolor{LightGray}{gray}{0.95}
\usepackage{float}
\usepackage{wrapfig}
\usepackage{overpic}
\usepackage{stfloats}

\ifCLASSOPTIONcompsoc
  \usepackage[nocompress]{cite}
\else
  \usepackage{cite}
\fi

\ifCLASSINFOpdf
\else
\fi

\def\ie{\emph{i.e.}}

\def\wrt{\emph{w.r.t.}}
\def\etal{\emph{et al.}}

\def\ie{\emph{i.e.}}

\def\wrt{\emph{w.r.t.}}
\def\etal{\emph{et al.}}

\usepackage{booktabs}
\usepackage{multirow}
\usepackage{tabularx}
\usepackage{wasysym}
\usepackage{colortbl}
\usepackage{xcolor}
\usepackage{hhline}
\usepackage{graphicx}
\usepackage{xcolor}
\definecolor{deepgreen}{RGB}{0,100,0}

\usepackage{ragged2e}
\usepackage[edges]{forest}
\usetikzlibrary{shadows.blur}
\usetikzlibrary{shapes.geometric}
\definecolor{hiddendraw}{RGB}{10,128,122}
\definecolor{hidden-orange}{RGB}{224,224,224}
\usepackage{subcaption}
\usepackage{svg}

\usepackage{pifont}
\usepackage[perpage,symbol*]{footmisc}
\DefineFNsymbols{circled}{{\ding{192}}{\ding{193}}{\ding{194}}
{\ding{195}}{\ding{196}}{\ding{197}}{\ding{198}}{\ding{199}}{\ding{200}}{\ding{201}}}
\setfnsymbol{circled}

\newcommand\myfootnotestyle[1]{\ifcase#1 \or \ding{182}\or \ding{183}\or
\ding{184}\or \ding{185}\or \ding{186}\or \ding{187}%
\or \ding{188}\or \ding{189}\or \ding{190}\or \ding{191}\else *\fi\relax}

\hypersetup{hidelinks,colorlinks=true}

\newif\ifsubmit
\submitfalse

\def\vtheta{{\bm{\theta}}}

\def\vx{{\bm{x}}}

\def\gL{{\mathcal{L}}}

\def\gW{{\mathcal{W}}}
\def\gX{{\mathcal{X}}}
\def\gY{{\mathcal{Y}}}

\def\w{\bm{w}}
\def\x{\bm{x}}

\def\vtheta{\bm{\theta}}
\def\vdelta{\bm{\delta}}

\DeclareMathOperator*{\argmax}{arg\,max}
\DeclareMathOperator*{\argmin}{arg\,min}

\usepackage[utf8]{inputenc} % allow utf-8 input
\usepackage[T1]{fontenc}    % use 8-bit T1 fonts
\usepackage{url}            % simple URL typesetting
\usepackage{booktabs}       % professional-quality tables
\usepackage{amsfonts}       % blackboard math symbols
\usepackage{nicefrac}       % compact symbols for 1/2, etc.
\usepackage{microtype}      % microtypography
\usepackage{xcolor}         % colors
\usepackage{amssymb}
\usepackage{float}
\usepackage{amsmath}
\usepackage{amsthm}
\usepackage{graphicx}
\usepackage{diagbox}
\usepackage{tabularx}
\usepackage{bm}
\usepackage{bbm}
\usepackage{algorithm}
\usepackage{algorithmic}
\usepackage{setspace}

\usepackage{amsfonts}
\usepackage{float}
\usepackage{enumitem,graphicx}
\usepackage{wrapfig}
\usepackage{multirow}
\usepackage{xcolor}
\def\w{\bm{w}}
\def\x{\bm{x}}

\def\vtheta{\bm{\theta}}
\def\vdelta{\bm{\delta}}

\def\vTheta{\bm{\Theta}}

\def\gL{\mathcal{L}}
\makeatletter
\DeclareRobustCommand\onedot{\futurelet\@let@token\@onedot}
\def\@onedot{\ifx\@let@token.\else.\null\fi\xspace}

\def\ie{\emph{i.e.}}

\def\wrt{\emph{w.r.t.}} 

\def\etal{\emph{et al. }}

\long\def\comment#1{}

\usepackage{epsfig}
\usepackage{graphicx}
\usepackage{amsmath}
\usepackage{amssymb}
\usepackage{dsfont}
\usepackage{xcolor}

\newtheorem{thm}{Theorem}
\begin{document}

\title{Class-Conditional Neural Polarizer: A Lightweight and Effective Backdoor Defense by Purifying Poisoned Features}

\author{%
  Mingli Zhu\footnote[1]{}, Shaokui Wei, Hongyuan Zha, Baoyuan Wu \textit{Senior Member, IEEE}
\thanks{M. Zhu, S. Wei, and B. Wu are with the School of Data Science, The Chinese University of Hong Kong, Shenzhen, Guangdong, 518172, P.R. China. (Email: minglizhu@link.cuhk.edu.cn, shaokuiwei@link.cuhk.edu.cn, wubaoyuan@cuhk.edu.cn)}
\thanks{H. Zha is with the School of Data Science, The Chinese University of Hong Kong, Shenzhen, and the Shenzhen Key Laboratory of Crowd Intelligence Empowered Low-Carbon Energy Network, China. (Email: zhahy@cuhk.edu.cn)}
\thanks{Corresponding author: Baoyuan Wu (wubaoyuan@cuhk.edu.cn).}
\thanks{M. Zhu and S. Wei contributed equally to this work.}
}

\markboth{Submitted to IEEE TRANSACTIONS ON PATTERN ANALYSIS AND MACHINE INTELLIGENCE}%
{Shell \MakeLowercase{\textit{\etal}}: Bare Demo of IEEEtran.cls for Computer Society Journals}


\maketitle

\begin{abstract}

Recent studies have highlighted the vulnerability of deep neural networks to backdoor attacks, where models are manipulated to rely on embedded triggers within poisoned samples, despite the presence of both benign and trigger information. While several defense methods have been proposed, they often struggle to balance backdoor mitigation with maintaining benign performance.
In this work, inspired by the concept of optical polarizer—which allows light waves of specific polarizations to pass while filtering others—we propose a lightweight backdoor defense approach, \textit{NPD}. This method integrates a \textbf{neural polarizer} (NP) as an intermediate layer within the compromised model, implemented as a lightweight linear transformation optimized via bi-level optimization. The learnable NP filters trigger information from poisoned samples while preserving benign content. 
Despite its effectiveness, we identify through empirical studies that NPD's performance degrades when the target labels (required for purification) are inaccurately estimated. 
To address this limitation while harnessing the potential of targeted adversarial mitigation, we propose class-conditional neural polarizer-based defense (\textit{CNPD}). 
The key innovation is a fusion module that integrates the backdoored model's predicted label with the features to be purified. This architecture inherently mimics targeted adversarial defense mechanisms without requiring label estimation used in NPD.
We propose three implementations of CNPD: the first is \textit{r-CNPD}, which trains a replicated NP layer for each class and, during inference, selects the appropriate NP layer for defense based on the predicted class from the backdoored model. To efficiently handle a large number of classes, two variants are designed: \textit{e-CNPD}, which embeds class information as additional features, and \textit{a-CNPD}, which directs network attention using class information.
Additionally, we provide a theoretical guarantee for the feasibility of the proposed CNPD method, and demonstrate that CNPD establishes an upper bound on backdoor risk.
Extensive experiments show that our lightweight and effective methods outperform existing methods across various neural network architectures and datasets.
\end{abstract}


\section{Introduction\label{sec1}}

\IEEEPARstart{D}{eep} Neural Networks (DNNs) have achieved remarkable performance in lots of critical domains, such as facial recognition \cite{sharif2016accessorize} and autonomous driving \cite{yurtsever2020survey}.
However, the susceptibility of DNNs to deliberate threats \cite{dong2021query,goldblum2022dataset,yin2023generalizable,wu2023adversarial} poses a serious challenge, as it imperils the integrity of these systems and undermines their reliability. Notably, the risk posed by backdoor attacks is garnering increased attention due to their inherent sophistication and stealthiness \cite{wu2024backdoorbench,zhu2024breaking}.
Backdoor attacks entail the malicious manipulation of the training dataset \cite{gu2019badnets} or training process \cite{liang2024badclip}. This results in a backdoored model which classifies poisoned samples with \textit{triggers} to a predefined \textit{target label} and behaves normally on benign samples \cite{gu2019badnets}. Backdoor attacks may originate from a multitude of avenues, such as the use of contaminated training datasets, employing third-party platforms for model training, or procuring pre-trained models from untrusted sources. These circumstances considerably heighten the risk posed by backdoor attacks on DNN's applications. Concurrently, they underscore the cruciality of devising backdoor defenses against such attacks.

This work focuses on \textbf{post-training defense}, which aims to mitigate the backdoor effect in a backdoored model, using a small subset of benign data, while preserving the model's performance on benign inputs.
Current approaches in this category mainly involve global fine-tuning \cite{zeng2022adversarial,zhu2023enhancing} or network pruning \cite{liu2018fine,zheng2022pre}. However, these methods face two challenges. First, the large parameter space involved in fine-tuning or pruning makes it difficult for a limited amount of benign data to effectively remove backdoors without compromising model performance. Second, the optimization of this parameter space is computationally expensive.

\begin{figure}
\centering
\vspace{0.5em}
\includegraphics[width=0.49\textwidth]{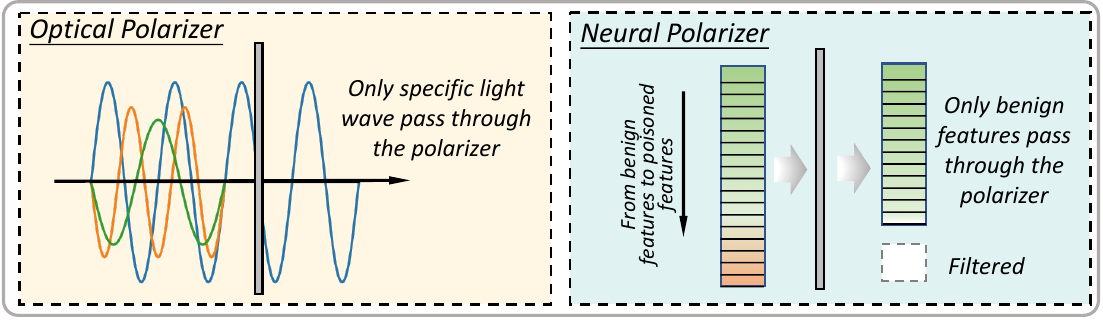}
\caption{Comparison of Optical and Neural Polarizers. An optical polarizer allows only light waves with specific polarizations to pass through. Similarly, in a neural polarizer integrated into a compromised model, only benign features are allowed to pass, while backdoor-related features are filtered out, effectively removing the backdoor.}
\label{motivation}
% \vspace{-4.15em}
\end{figure}

To address the above constraints, we introduce a lightweight yet effective countermeasure that operates without requiring model retraining. 
Our approach is inspired by optical polarizers \cite{xiong2016optical}, which allow only light waves with specific polarizations to pass while blocking others in a mixed light wave (refer to Fig. \ref{motivation}). 
Similarly, our method involves inserting one learnable layer while keeping all layers of the original backdoored model fixed. By applying a similar principle, if we consider a poisoned sample as a combination of both trigger and benign features, we can design a \textit{neural polarizer} (NP) to filter the trigger features while conserving the benign ones. Consequently, contaminated samples could be cleansed thereby diminishing the impact of the backdoor, while benign samples remain largely unaffected.
With the lightweight NP, our objective is to weaken the correlation between the trigger and the target label, where both of them are inaccessible to the defender. To tackle it, we first formulate a bi-level optimization problem, where in inner optimization, we dynamically estimate the target label and trigger with adversarial examples (AEs), and the NP is updated in outer minimization by unlearning these AEs. We term this method as Neural Polarizer based backdoor Defense (\textit{NPD}).

In this process, we observe that the effectiveness of defense depends on the accuracy of the estimated target label. An incorrect estimation can either fail to remove the backdoors or cause oscillations in the attack success rate during training. To address this limitation, we observe that the target label is available when an attacker successfully launches an attack. Therefore, we explore leveraging the output of the backdoored model as a proxy for the target label, utilizing the ``home field'' advantage of defenders. To this end, we propose class-conditional neural polarizer-based backdoor defense (\textit{CNPD}). This method introduces a fusion module that combines class information predicted by the backdoored model with the features to be purified, effectively emulating the targeted adversarial approach. We propose three implementations of CNPD. The first, replicated Class-conditional NPD (r-CNPD),  which trains a replicated NP layer for each class in post-training phase, and in inference phase, the class predicted by the backdoored model is used to select the corresponding NP layer for feature purification. This approach demonstrates performance comparable to the targeted adversarial defense method.
Given the computational cost increases with the number of classes, two variants are designed to mitigate this complexity. The first, \textit{e-CNPD}, integrates class embeddings as additional features to assist in feature purification. The second, \textit{a-CNPD}, uses class-related information to guide the model's attention toward the features requiring purification, inspired by attention mechanisms \cite{vaswani2017attention}. These approaches enable efficient backdoor purification while keeping computational complexity manageable. Additionally, we provide a theoretical guarantee for the feasibility of CNPD, and demonstrate that it offers an upper bound on the backdoor risk \cite{wei2023shared}.

Our main contributions are fourfold:
1) We propose a novel backdoor defense paradigm, NPD, that optimizes only one additional layer while keeping the original parameters of the backdoored model fixed.
2) Based on NPD, we introduce the Class-conditional NPD (CNPD), which uses the backdoored model to guide the purification of poisoned samples. We design three variations-\textit{r-CNPD}, \textit{e-CNPD}, and \textit{a-CNPD}-to implement this approach.
3) We provide a theoretical analysis to validate the feasibility of our CNPD approach, and establish an upper bound on backdoor risk.
4) We conduct extensive experiments and analyses, including comparisons with state-of-the-art (SOTA) methods, and visualization analysis.

This paper extends our conference paper (\textit{NPD} \cite{zhu2024neural}) with significant improvements. Compared to the conference version, the current version introduces the following key enhancements:
1) We introduce CNPD, which leverages the backdoored model to guide poisoned sample purification.  
2) We propose \textit{r-CNPD} and introduce two additional efficiency-boosting mechanisms—\textit{e-CNPD}  and \textit{a-CNPD} —which embed label information to more effectively filter out backdoor features.
3) We provide a theoretical analysis to validate CNPD’s feasibility and show that it offers an upper bound on backdoor risk compared to the original NPD.  
4) We conduct additional experiments, including comparisons with SOTA methods, ablation studies, and visualization analyses.

% The rest of the paper is organized as follows: Section \ref{sec2} reviews the development trends in backdoor attacks, defenses, and detections. Section \ref{sec3} provides a detailed introduction to the three proposed methods, with theoretical analysis. Section \ref{sec4} presents the results of extensive experiments and analysis. Section \ref{sec5} concludes the paper, and we hope this work contributes to enhancing the robustness of DNNs against backdoor attacks and inspiring further research into more efficient and practical defense strategies.

\section{Related work\label{sec2}}
\subsection{Backdoor attacks}
Deep neural networks (DNNs) have been found to be vulnerable to backdoor attacks, where the backdoored model performs normally on benign inputs while predicting a specified target label when encounters input with the pre-defined trigger. Backdoor attacks are generated by poisoning training dataset \cite{gu2019badnets,chen2017targeted,zeng2021rethinking,li2021invisible} or injecting backdoors in the training process \cite{nguyen2020input,nguyen2021wanet,liang2024badclip,foret2021sharpnessaware}.
In dataset poisoning attacks, the adversary maliciously modifies the dataset by introducing carefully crafted triggers onto images and altering the corresponding labels to a target label. As a result, the backdoored model learns incorrect associations between the triggers and the intended target label.
The seminal work by Gu \etal \cite{gu2019badnets} introduced the "BadNets", demonstrating that subtle alterations to a fraction of training samples could lead to a targeted misclassification during inference. This foundational study spurred extensive exploration into various mechanisms of backdoor insertion, particularly focusing on invisible triggers design \cite{li2021invisible,zeng2021rethinking}, dataset poisoning with low poisoning ratio \cite{song2024wpda,zhu2023boosting}, or clean-label attacks \cite{gao2021clean,shafahi2018poison,barni2019new} that crafts poisoned samples without modifying labels. These works increase the stealthiness of backdoor attacks, and highlight the necessity for improved detection and mitigation techniques. 

In training controllable backdoor attacks, the adversary strategically optimizes the triggers and backdoored models meanwhile to enhance the stealthiness and impact of the triggers, ensuring they can be effectively activated by the model. Additionally, the adversary seeks to make the model more resilient to backdoor detection and mitigation efforts.
For instance, WaNet \cite{nguyen2021wanet} proposes a novel "noise" training mode to increase the difficulty of detection. He \etal \cite{he2024sharpnessaware} proposes sharpness-aware data poisoning
attack that maintains the poisoning effect under various re-training process. Such advancements underscore the increasing sophistication of backdoor strategies, emphasizing the growing need for comprehensive defense method in AI systems.
Beyond the traditional images domain, backdoor attacks have also proven to be highly effective against self-supervised models \cite{jia2022badencoder,li2023embarrassingly}, diffusion models \cite{chou2024villandiffusion,chou2023backdoor}, large language models \cite{yang2024watch,li2024backdoorllm}, and multimodal models \cite{liang2024badclip,bai2024badclip}, posing a significant threat to deep learning.

\subsection{Backdoor defense}
Numerous studies \cite{wu2023defenses,niu2024towards} have been conducted to develop defenses against the threats posed by backdoor attacks. According to defense stages, existing studies can be roughly partitioned into four categories: pre-processing \cite{chen2022effective,zhu2024vdc}, in-training \cite{li2021anti,huang2022backdoor}, post-training \cite{wang2019neural,li2021neural}, and inference stage defenses \cite{guoscale,chou2020sentinet}. 
Pre-processing defenses aim to remove poisoned samples from a compromised dataset by input anomaly detection. 
Training-stage strategies assume that a defender is given the poisoned training dataset and the defender needs to train a secure model and inhibit backdoor injection based on the poisoned dataset. These strategies exploit differences between poisoned and clean samples to filter out potentially harmful instances \cite{chen2022effective}. 
Post-training defenses assume that a defender has access to a backdoored model and a limited number of clean samples. These defenses aim to eliminate backdoor effects from compromised models without requiring knowledge of the trigger or target labels. Techniques include pruning potentially compromised neurons \cite{liu2018fine,zheng2022pre} and performing strategic fine-tuning \cite{zeng2022adversarial,zhu2023enhancing}. For fine-tuning based methods, I-BAU \cite{zeng2022adversarial} establishes a minimax optimization framework for training the network using samples with universal adversarial perturbations. Similarly, PBE \cite{mu2023progressive} leverages untargeted adversarial attacks to purify backdoor models. SAU \cite{wei2023shared} firstly searches for shared adversarial example between the backdoored model and the purified model and unlearns the generated adversarial example for backdoor purification. 
\textit{These studies leverage adversarial examples in backdoor models to mitigate backdoor threats; however, none have discussed the impact of different adversarial examples on the effectiveness of backdoor removal.}
Inference stage defenses seek to prevent backdoor activation through sample detection and destroying trigger's feature. 
In this work, we mainly consider post-training backdoor defense, and follow their default settings.

\section{Methodology\label{sec3}}
In this section, we present the details of our methodology. First, in Sec. \ref{sec3.1}, we define the basic settings, including the notations and the threat model.  
Next, in Sec. \ref{sec3.2}, we introduce our Neural Polarizer-based Defense (NPD) method, outlining its objective, optimization, and algorithm.  
In Sec. \ref{sec3.3}, we analyze the limitation of NPD and present our Class-conditional NPD method, followed by the introduction of three distinct implementations in Sec. \ref{sec3.4}.  
Finally, in Sec. \ref{sec3.6}, we provide a theoretical analysis to support the feasibility and effectiveness of our approach.

\subsection{Basic Settings\label{sec3.1}}
\subsubsection{Notations} We consider a classification problem with $C$ classes ($C\geq 2$). Assume $\x\in \gX \subset \mathbb{R}^d$ to be a $d$-dimensional input sample, its corresponding ground truth label is denoted as $y\in \gY = \{1,\dots,C\}$. A deep neural network comprising of $L$ layers $f:\gX\times\gW\to \mathbb{R}^C$, parameterized by $\w\in\gW$, is defined as follows:
\begin{equation}
f(\x ; \w)=f^{(L)}_{\w_{L}} \circ f^{(L-1)}_{\w_{L-1}} \circ \cdots \circ f^{(1)}_{\w_{1}} (\x),
\end{equation}
where $f^{(l)}_{\w_{l}}$ embodies the function (e.g., convolution layer or batch normalization layer) with parameter $\w_{l}$ located in the $l^{\text {th }}$ layer, where $1 \leq l \leq L$. To maintain simplicity, $f(\x ; \w)$ is presented as $f_{\w}(\x)$ or $f_{\w}$. Given an input $\x$, the predicted label of $\x$ is determined by $\argmax_c f_c(\x;\w), c=1,\ldots,C$, where $f_c(\x; \w)$ represents the logit of the $c^{\text {th}}$ class.

\subsubsection{Threat model} 
\textbf{Attacker's goal.} In this work, we consider the scenario that the attacker is an unauthenticated third-party model publisher or dataset publisher. Thus the backdoored model $f_{\w}$ can be generated through the manipulation of either the training process or the training data, ensuring that $f_{\w}$ functions correctly on benign sample $\x$ (\ie, $f_{\w}(\x) = y$), and predicts the poisoned sample $\x_{\Delta} = r(\x, \Delta)$ to the targeted label $T$. Here, $\Delta$ indicates the pre-defined trigger and $r(\cdot, \cdot)$ is the fusion function that integrates $\x$ and $\Delta$. 
Considering the possibility of the attacker establishing multiple target labels, we assign $T_i$ to represent the target label for sample $\x_i$.

This threat is practical in the real world applications that heavily rely on AI-driven decisions, particularly in scenarios where organizations use third-party models or datasets without thoroughly scrutinizing their integrity due to resources limits.
The stealthiness of backdoor attacks makes them particularly hard to immediately recognize, thus posing substantial potential risks for individuals or entities.

\noindent \textbf{Defender's capabilities and objectives.}
Assume that the defender is given the backdoored model $f_{\w}$ and has access to a limited number of benign training samples, denoted as $\mathcal{D}_{bn}=\{(\x_i,y_i)\}_{i=1}^{N}$. 
The defender's goal is to obtain a new model $\hat{f}$ based on $f_{\w}$ and $\mathcal{D}_{bn}$, such that the backdoor effect is removed and the benign performance is maintained in $\hat{f}$, \ie, $\hat{f}(\x_{\Delta}) \neq y$ and $\hat{f}(\x) = y$. Note that the defender is unaware of the attacker's target labels and specified triggers, and can only leverage a limited number of samples to modify the backdoored model. This presents a stringent yet highly practical scenario; for instance, a user may lack the resources to train a model from scratch and instead obtain a model from a third-party unauthenticated organization. However, fine-tuning the model is affordable to achieve a secure model.

\begin{figure*}
\centering
\vspace{0.5em}
\includegraphics[width=\textwidth]{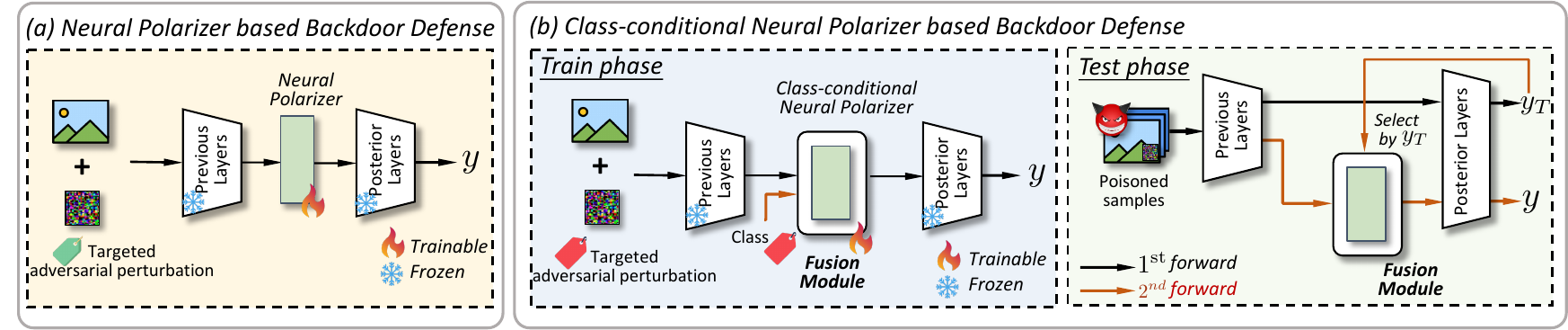}
\caption{\textbf{(a)}: Neural Polarizer based Backdoor Defense. Backdoor defense by integrating a trainable neural polarizer into the compromised model.
\textbf{(b)}: Class-conditional Neural Polarizer based Backdoor Defense. During training, a trainable neural polarizer is incorporated into the compromised model, with a fusion module that fuses internal features and class information. During inference, the output of the backdoored model is used to guide class-conditional neural polarizer for feature filtering.}
\label{two_methods}
% \vspace{-4.15em}
\end{figure*}

\subsection{Neural Polarizer based Backdoor Defense\label{sec3.2}}
We propose the neural polarizer (NP) to purify poisoned samples in the feature space. As shown in Fig. \ref{two_methods} \textbf{(a)}, the neural polarizer $g_{\vtheta}$ is inserted into the backdoored model $f_{\w}$ at a specific immediate layer, to obtain a combined network $f_{\w, \vtheta}$, \ie, $f_{\w, \vtheta}=f^{(L)}_{\w_{L}} \circ \cdots \circ f^{(l+1)}_{\w_{l+1}} \circ g_{\vtheta} \circ f^{(l)}_{\w_{l}} \circ \cdots \circ f^{(1)}_{\w_{1}} (\x)$. For clarity, we denote $f_{\w, \vtheta}$ as $\hat{f}_{\vtheta}$, since $\w$ is fixed.

\subsubsection{Objective of the neural polarizer\label{sec3.2.1}}
Informally, a desired neural polarizer should have the following three properties:
\begin{itemize}[leftmargin=*]
    \item \textbf{Compatible with neighboring layers}: This implies that both the input feature and output activation should be of the same shape, which can be achieved through careful architectural design.
    \item \textbf{Filtering trigger features in poisoned samples}: After applying the neural polarizer, the trigger features should be filtered out to ensure that the backdoor is deactivated, \ie, $\hat{f}_{\vtheta}(\x_{\Delta})\neq T$.
    \item \textbf{Preserving benign features in both poisoned and benign samples}: The neural polarizer should maintain benign features to ensure that $\hat{f}_{\vtheta}$ performs effectively on both poisoned and benign samples, \ie, $\hat{f}_{\vtheta}(\x_{\Delta})= \hat{f}_{\vtheta}(\x) = y$.     
\end{itemize}
The first property could be easily satisfied by carefully designing the architecture of the neural polarizer. For example, given the input feature map \(m^l(\x) \in \mathbb{R}^{C_l \times H_l \times W_l}\), the neural polarizer is implemented by a convolution layer (Conv) with $C_l$ convolution filters of shape $1\times1$, followed by a Batch Normalization (BN) layer. The Conv-BN block can be seen as a linear transformation layer. 
To satisfy the latter two properties, the parameters $\vtheta$ should be learned by solving a carefully formulated optimization problem as presented in Sec. \ref{sec3.2.2}.

\subsubsection{Learning neural polarizer\label{sec3.2.2}}
\textbf{Loss functions in oracle setting.}
To learn a desired neural polarizer $g_{\vtheta}$, we begin with an oracle scenario in which the trigger $\Delta$ and the target label $T$ are provided. We introduce several loss functions designed to encourage the NP $g_{\vtheta}$ to fulfill the two properties previously mentioned, as outlined below:
\begin{itemize}[leftmargin=*]
    \item \textbf{Loss for filtering trigger features in poisoned samples.} Given trigger $\Delta$ and target label $T$, trigger features can be filtered  by weakening the connection between \(\Delta\) and \(T\) using the following loss function:
    \begin{equation}
        \gL_{asr}(\x,\Delta, T; \vtheta) = -\log(1 - s_{T}(\x_{\Delta}; \vtheta)),
        \label{eq:asr_loss}
    \end{equation}
    where \(s_{T}(\x_{\Delta}; \vtheta)\) represents the softmax probability of the model \(\hat{f}_{\vtheta}\) predicting \(\x_{\Delta}\) as label \(T\).
    \item \textbf{Loss for maintaining benign features in poisoned samples.} To ensure that a poisoned sample can be correctly classified to its true label, we utilize the boosted cross-entropy loss introduced in \cite{wang2020improving}:
    \begin{equation}
    \begin{aligned}
        \gL_{bce}(\x, y, \Delta; \vtheta) &= -\log(s_{y}(\x_{\Delta}; \vtheta)) \\
        &\quad - \log\big(1 - \max_{k \neq y} s_{k}(\x_{\Delta}; \vtheta)\big).
        \label{eq:bce_loss}
    \end{aligned}
    \end{equation}
    \item \textbf{Loss for maintaining benign features in benign samples.} To preserve benign features of benign samples, the standard cross-entropy loss is employed:
    \begin{equation}
        \gL_{bn}(\x, y; \vtheta) = -\log(s_{y}(\x; \vtheta)).
    \end{equation}
\end{itemize}
% \
% \newline
% \noindent
% \quad 
% \textbf{Approximating \(T\) and $\Delta$.}
\noindent \textbf{Approximating \(T\) and $\Delta$.}
Note that the target label \(T\) and trigger \(\Delta\) is inaccessible in both \(\gL_{asr}\) and \(\gL_{bce}\). As a result, these two losses cannot be directly optimized and require an approximation. In terms of approximating \(T\),  although several methods \cite{ma2022beatrix,guo2021aeva} have been developed to detect the target label of backdoored models, we adopt a straightforward, sample-specific, and dynamic strategy: 
\begin{equation}
    T\approx y' = \argmax_{k\neq y} \hat{f}_k(\x;\vtheta).
    \label{eq:label}
\end{equation}
This strategy offers two key advantages. First, the predicted target label allows us to generate targeted adversarial perturbations. As discussed later, targeted perturbations serve as a better surrogate for the unknown trigger compared to untargeted ones. Second, the sample-specific target label prediction is versatile and works for both all2one (one trigger, one target class) and all2all (every class as a potential target) attack settings. In practice, defenders often lack certainty regarding the attack type—whether it’s all2one, all2all, or multi-trigger multi-target—leading to potential sub-optimal defense strategies due to incorrect assumptions or detection failures. Our approach, with its sample-specific target label prediction, mitigates this risk.

In terms of approximating \(\Delta\), it is dynamically approximated by the targeted adversarial perturbation (T-AP) \cite{madry2017towards} of \(\hat{f}_{\vtheta}\), \ie, 

\begin{equation}\label{eq:at}
\Delta \approx \vdelta^* = \argmin_{\|\vdelta\|_p\leq \rho} \gL_{\text{CE}}\left(\hat{f}_{\vtheta}(\x+\vdelta), y'\right),
\end{equation}
where $\|\cdot\|_p$ is the $L_p$ norm, $\rho$ is the budget for perturbations.

\noindent \textbf{Bi-level optimization problem.}
By utilizing the approximation of \(T\) and \(\Delta\) as shown in Eq. (\ref{eq:label}) and (\ref{eq:at}), and substituting them into the loss functions in Eq. (\ref{eq:asr_loss}) and Eq. (\ref{eq:bce_loss}), we formulate the following optimization problem to learn \(\vtheta\) based on the clean dataset \(\mathcal{D}_{bn}\):
\begin{equation}\label{loss}
\begin{aligned}
    \min_{\vtheta} \quad \quad & \frac{1}{N}\sum_{i=1}^N \lambda_1 \gL_{bn}({\x_i,y_i};\vtheta) +  \lambda_2 \gL_{asr}({\x_i,y_i,\vdelta^*_i, y'_i;\vtheta}) \\
    & + \lambda_3 \gL_{bce}({\x_i,y_i,\vdelta^*_i;\vtheta}),\\
    \mathrm{s.t.} \quad \quad & \vdelta^*_i = \argmin_{\|\vdelta_i\|_p\leq \rho} \gL_{\text{CE}}\left(\hat{f}_{\vtheta}(\x_i+\vdelta_i), y'_i\right), \\
    & y'_i = \argmax_{k_i \neq y_i} \hat{f}_{k_i}(\x_i;\vtheta),
\end{aligned}
\end{equation}
where $\lambda_1,\lambda_2,\lambda_3 > 0 $ are hyper-parameters to adjust the importance of each loss function. 
This bi-level optimization is dubbed Neural Polarizer based backdoor Defense (\textbf{NPD}).
NPD is highly efficient, as the lightweight NP layer contains only a few parameters that need optimization, allowing it to be learned with a limited number of clean samples.

\ 
\newline
\noindent
\quad 
\textbf{Optimization algorithm.} We propose Algorithm \ref{alg1_npd} to solve the above NPD. Specifically, the key steps involve alternatively updating the perturbation $\vdelta$ and $\vtheta$ as follows:
\begin{itemize}[leftmargin=*]
    \item \textbf{Inner minimization:} Given parameters $\vtheta$, estimate the target label $y'$ by Eq. (\ref{eq:label}). Then the targeted Project Gradient Descent (PGD) \cite{madry2017towards} is employed to generate the perturbation $\vdelta^*$ via Eq. (\ref{eq:at}).
    \item  \textbf{Outer minimization:} Given $y'$ and $\vdelta^*$ for each sample in a batch, the $\vtheta$ can be updated by taking one stochastic gradient descent \cite{bottou2007tradeoffs} (SGD) step \wrt~ the outer minimization objective in Eq. ($\ref{loss}$).
\end{itemize}

\begin{algorithm}[h]
\caption{Neural Polarizer based Backdoor Defense}\label{alg1_npd}
\begin{algorithmic}[1]
\STATE \textbf{Input:} Training set $\mathcal{D}_{bn}$, backdoored model $f_{\w}$, neural polarizer $g_{\vtheta}$, learning rate $\eta>0$, perturbation bound $\rho>0$, norm $p$, hyper-parameters $\lambda_1$,$\lambda_2$,$\lambda_3>0$, warm-up epochs $\mathcal{T}_0$, training epochs $\mathcal{T}$, number of PGD steps $T_{adv}$.
\vspace{-4mm}
\STATE \textbf{Output:} Model $\hat{f}(\w,\vtheta)$.
\STATE Initialize $g_{\vtheta}$ to be an identity function. Fix $\w$, and construct the composed network $\hat{f}(\w,\vtheta)$.
\STATE Warm-up: Train $g_{\vtheta}$ with $\mathcal{L}_{bn}$ for $\mathcal{T}_0$ epochs.
\FOR {$t=0,...,$ $\mathcal{T} -1$}
\FOR {mini-batch $\mathcal{B}=\{(\x_i,y_i)\}_{i=1}^{b}\subset \mathcal{D}_{bn}$}
\STATE Compute $\{y'_i\}_{i=1}^{b}$ by Eq. (\ref{eq:label}); 
\STATE Generate perturbations $\{(\boldsymbol{\delta_i})\}_{i=1}^{b}$ with $\boldsymbol{\|\delta_{i}\|_p}\leq \rho$ and target labels $\{y'_i\}_{i=1}^{b}$ by targeted PGD attack \cite{madry2017towards} via Eq. (\ref{eq:at});
\STATE Update $\vtheta$ via outer minimization of Eq. (\ref{loss}) by SGD.
\ENDFOR
\ENDFOR
\RETURN Model $\hat{f}(\w,\vtheta)$. 
\end{algorithmic}
\end{algorithm}

\subsection{Class-conditional NPD\label{sec3.3}}
\subsubsection{Limitations of NPD\label{sec3.3.1}}
\begin{figure}[h]
\centering
\vspace{0.5em}
\includegraphics[width=0.4\textwidth]{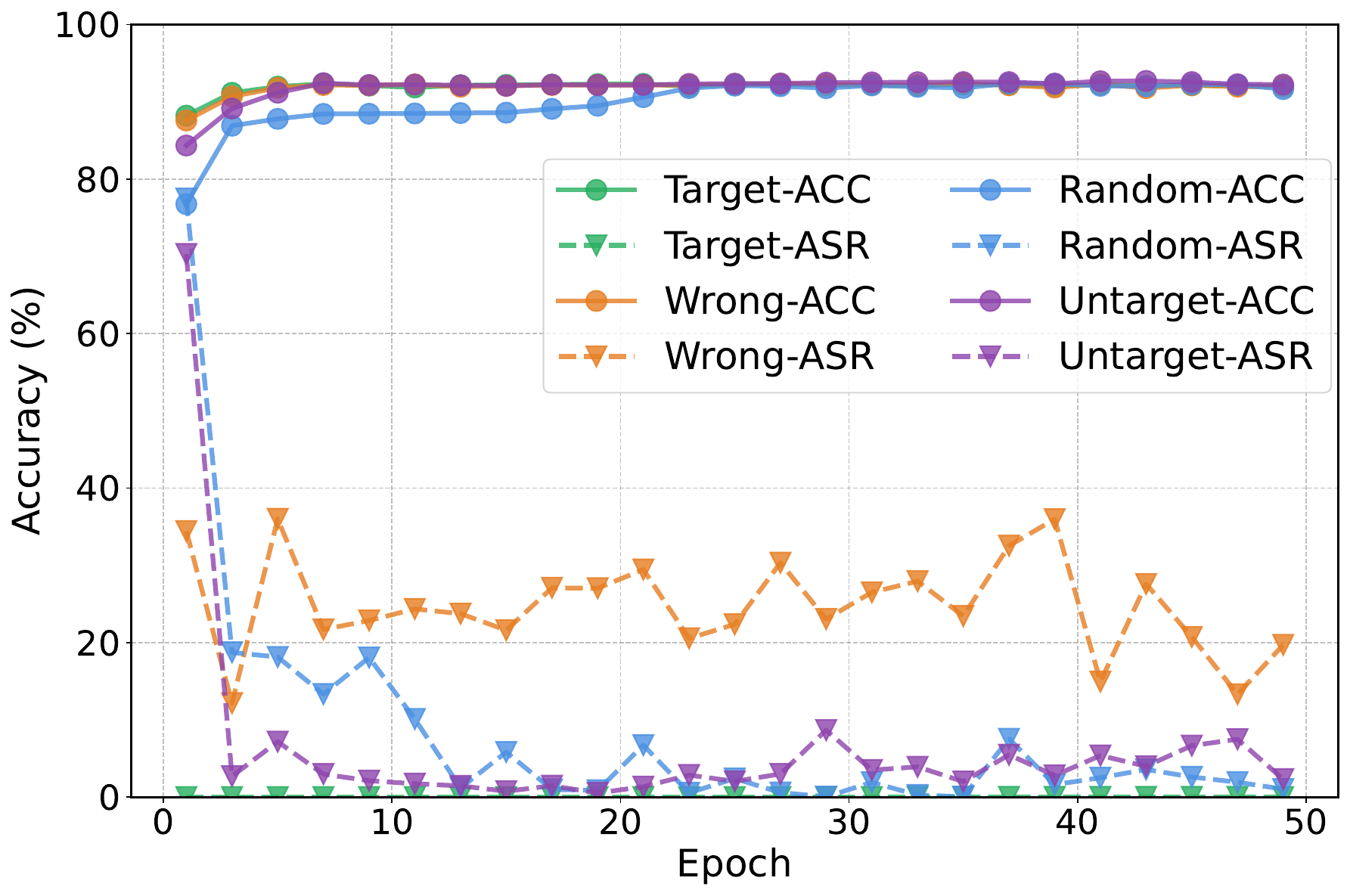}
\caption{Defense performance of unlearning adversarial examples (AEs) using different strategies on Trojan attack \cite{Trojannn}. ``Target'' refers to unlearning AEs that use the attacker's target label. ``Wrong'' indicates unlearning AEs assigned a label that is not the attacker's target label. ``Random'' signifies unlearning AEs with randomly assigned labels. ``Untarget'' represents unlearning AEs with an untargeted objective.}
\label{moti}
% \vspace{-4.15em}
\end{figure}

One limitation of NPD lies in its reliance on the accuracy of the estimated target label (see Eq. (\ref{eq:label})). An inaccurate estimation may either fail to eliminate backdoors or result in oscillations in the attack success rate during training.
To further investigate this, we conduct experiments to compare the defense performance using different AEs. 
In detail, we use the same model architecture and bi-level optimization framework as in NPD, and substitute the AEs generation method with four different AEs generation methods. As shown in Fig. \ref{moti}, ``Target'' refers to unlearning targeted AEs (using the same target label as attacker's target label $T$), while ``Wrong'' involves unlearning AEs assigned a label different from the attacker's target. The ``Random'' consists of unlearning AEs with randomly assigned labels, and ``Untarget'' represents unlearning AEs with an untargeted objective. The experiment is conducted using the Trojan attack \cite{Trojannn} on the CIFAR-10 dataset with PreAct-ResNet18 network.
As the figure illustrates, the ``Wrong'' group fails in backdoor mitigation, and the ASR curves in the backdoor removal processes of the ``Random'' and ``Untarget'' groups exhibit oscillations. In contrast, the ``Target'' group removes the backdoor effectively and steadily with minimal impact on accuracy. This experiment highlights that different types of AEs result in significantly varying backdoor removal performance, with targeted AEs proving to be the most effective.

However, the target label remains unavailable to the defender in post-training stage. Fortunately, the target label is revealed when the attacker launches an attack during inference. Therefore, we propose utilizing the output of the backdoored model as a proxy for the target label, taking advantage of the defenders' "home field" advantage.

\subsubsection{Class-conditional NPD for DNNs\label{sec3.3.2}}
Unlike NPD, which unlearns targeted AEs based on estimated target labels, Class-conditional NPD (CNPD) designs a fusion module that is trained evenly across all classes, using class information as a condition for feature purification.
As dipicted in Fig. \ref{two_methods} \textbf{(b)}, the fusion module combines the class information \( y \) and the feature map $m^{l}(\x)$ as input, and the output of the module is:
\begin{equation}
    \hat{g}_{\vTheta}(m^{l}(\x),y) = S(g_{\vtheta}(m^{l}(\x)),y).
\end{equation}
Here, $S$ represents fusion operation, and $\vTheta$ is the parameters set of NP.
The combined network is denoted as $f_{\w, \vTheta}$ or $\hat{f}_{\vTheta}$, \ie, $f_{\w, \vTheta}=f^{(L)}_{\w_{L}} \circ \cdots \circ f^{(l+1)}_{\w_{l+1}} \circ \hat{g}_{\vTheta} \circ f^{(l)}_{\w_{l}} \circ \cdots \circ f^{(1)}_{\w_{1}} (\x)$.

The optimization of $\hat{g}_{\vTheta}$ is similar to the optimization of $g_{\vtheta}$ as introduced in Eq. (\ref{loss}). The only difference is that the second constraint in Eq. (\ref{loss}) becomes:
\begin{equation}
    y'_i \in \{1,\ldots,C\},i = 1,\ldots,N.
\end{equation}
It means the surrogate target label is equally distributed across all classes.
In the training phase, we train NP for each label evenly for backdoor mitigation. 
During the testing phase, it involves two forward passes. First, the predicted label $y'=\argmax f_{\vtheta}(\x)$ is obtained from the backdoored model $f_{\vtheta}$ without the NP. Then, the fusion module purifies $m^{l}(\x)$ based on $y'$ to produce the final output category.
\textit{This strategy makes sense because a successful attack must imply that $y'=T$, and thus the fusion module adjusts its parameters for defense.}

\begin{figure*}
\centering
\vspace{0.5em}
\includegraphics[width=\textwidth]{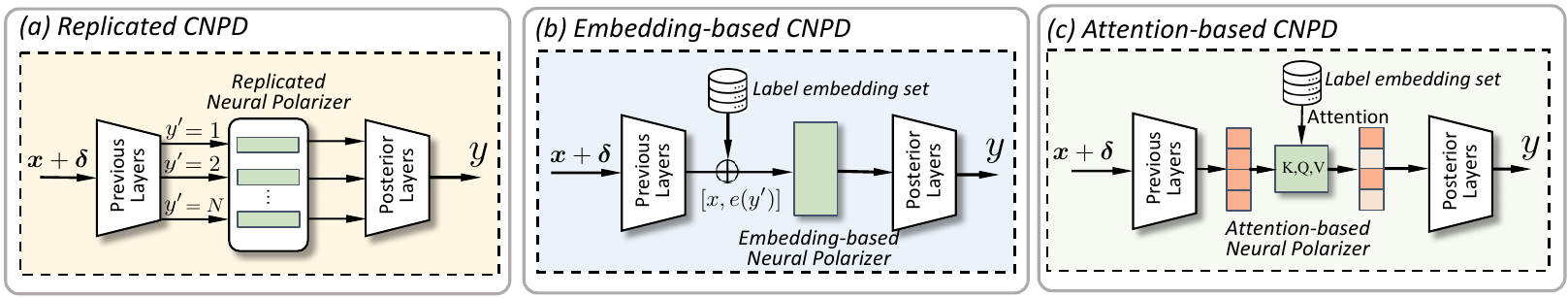}
\caption{
\textbf{(a)}. Replicated CNPD: Each class is associated with an individual neural polarizer.
\textbf{(b)}. Embedding-based CNPD: Class information is embedded as features within the model.
\textbf{(c)}. Attention-based CNPD: Class information is used to guide the network's attention for purification.
}
\label{three_structures}
% \vspace{-4.15em}
\end{figure*}
\subsection{Three Implementations of CNPD\label{sec3.4}}
To implement our CNPD approach, we design three distinct fusion module architectures, each accompanied by its corresponding training algorithm.

\subsubsection{Replicated CNPD\label{sec3.4.1}}
As dipicted in Fig. \ref{three_structures} \textbf{(a)}, the replicated CNPD (\textbf{r-CNPD}) defines a separate NP for each class $c$, parameterized by $\vtheta_c$, \ie, $\vTheta=\{\vtheta_c,c=1,\cdots,C\}$. Given an input feature map $m^{l}(\x)$ with label $y$, the output of the NP layer is:
\begin{equation}
    \hat{g}_{\vTheta}(m^{l}(\x),y) = g_{\vtheta_y}(m^{l}(\x)).
\end{equation}
In the training phase, the parameters $\vtheta_y$ are sequentially trained across all classes for backdoor mitigation. The training algorithm is shown in Alg. \ref{alg2_rnpd}.

\begin{algorithm}[h]
\caption{Replicated Class-conditional NPD (\textbf{r-CNPD})}\label{alg2_rnpd}
\begin{algorithmic}[1]
\STATE \textbf{Input:} Training set $\mathcal{D}_{bn}$, backdoored model $f_{\w}$, neural polarizer $\hat{g}_{\vTheta}$,
learning rate $\eta>0$, perturbation bound $\rho>0$, norm $p$, hyper-parameters $\lambda_1$,$\lambda_2$,$\lambda_3>0$, training epochs $\mathcal{T}$, number of PGD steps $T_{adv}$.
\STATE \textbf{Output:} Model $f(\w,\vTheta)$.
\STATE Initialize $g_{\vtheta_c}$ to be an identity function for $c\in {1,\ldots,C}$. Fix $\w$, and construct the composed network $f(\w,\vTheta)$.
\FOR {$c=1,...,C$}
\FOR {$t=0,...,$ $\mathcal{T} -1$}
\FOR {mini-batch $\mathcal{B}=\{(\x_i,y_i)\}_{i=1}^{b}\subset \mathcal{D}_{bn}$}
\STATE Generate perturbations $\{(\boldsymbol{\delta_i})\}_{i=1}^{b}$ with $\boldsymbol{\|\delta_{i}\|_p}\leq \rho$ and target label $c$ by targeted PGD attack \cite{madry2017towards} via Eq. (\ref{eq:at});
\STATE Update $\vtheta_c$ via outer minimization of Eq. (\ref{loss}) by SGD.
\ENDFOR
\ENDFOR
\ENDFOR
\RETURN Model $f(\w,\vTheta)$. 
\end{algorithmic}
\end{algorithm}

\subsubsection{Embedding-based CNPD\label{sec3.4.2}}
Although our r-CNPD method performs effectively in practice, the training cost and network capacity increase as the number of categories in the classification task grows, which may hinder its practical application. To address this issue, we design a fusion module that incorporates a fixed NP structure and a label embedding set, as illustrated in Fig. \ref{three_structures} (b).
Unlike r-CNPD, the \textbf{e-CNPD} method directly incorporates category information into the network as embedded features to influence the effect of different categories on feature filtering. Let the input feature be \(m^l(\x) \in \mathbb{R}^{C_l \times H_l \times W_l}\). We predefine specific embedded features for each category, denoted as \(e(c) \in \mathbb{R}^{1 \times H_l \times W_l}\), where \(c \in \{1, \ldots, C\}\). The concatenated feature \(\text{cat}[m^l(\x), e(y)]\) is then passed through the NP for filtering, defined as:
\begin{equation}
    \hat{g}_{\vTheta}(m^{l}(\x),y) = g_{\vTheta}(\text{cat}[m^{l}(\x),e(y)]).
\end{equation}
During each batch update in the training phase, we first randomly generate target labels for all batch samples and then create targeted adversarial samples to train the NP layer.

\subsubsection{Attention-based CNPD\label{sec3.4.3}}
Although the e-CNPD method is effective, careful consideration is required when designing the scale of feature fusion to avoid excessive distortion on the original features. To address this, we propose a more refined feature filtering method called Attention-based CNPD (\textbf{a-CNPD}). As shown in Fig. \ref{three_structures} (c), we design an attention-like \cite{vaswani2017attention} mechanism, enabling different labels to modulate the network's attention to features across channels. 
We denote the embedding of class \(c\) as \(e(c)\), where \(c \in \{1, \dots, C\}\), and the parameters of the two linear layers as \(\vtheta_Q\), \(\vtheta_K\), one convolutional layer as \(\vtheta_V\). The subsequent convolutional and batch normalization layers are denoted as \(g_{\vtheta_o}\). Given an input feature map \(m^l(\x)\) with label \(y\), the output of the network \(g_{\vtheta}\) is defined as:
\begin{equation}
\label{eq:anpd}
    \begin{aligned}
    	&\hat{g}_{\vTheta}(m^l(\x), y) = \\
    	&g_{\vtheta_o}(\text{softmax}\left[\frac{(\vtheta_Q \cdot e(y))^{\top} (\vtheta_K \cdot e(y))}{\sqrt{d_k}}\right]\cdot (\vtheta_V (m^l(\x)))),
    \end{aligned}
\end{equation}
where \(d_k\) is the dimension of \(\vtheta_K\).

The detailed algorithms for solving e-CNPD and a-CNPD are slightly different from those of r-CNPD, which are provided in Appendix A.

\subsection{Theoretical Analysis\label{sec3.6}}
In this section, we analyze the effectiveness of the CNPD method theoretically. 

\subsubsection{Existence of Class-conditional Neural Polarizer}
We first provide theoretical foundation for the existence of Class-conditional Neural Polarizer. To simplify our analysis, we focus on classification task equipped with a squared loss \cite{masnadi2008design} in a reproducing kernel Hilbert space $\mathcal{H}_{X}$ \cite{rosipal2001kernel} generated by kernel $\kappa_{X}$, feature mapping $\phi_{X}$ and inner product $\langle \cdot, \cdot\rangle_{\mathcal{H}_{X}}$. Then, for a function $h\in \mathcal{H}_{X}$, we have $h(\x)=\langle \phi_{X}(\x),h\rangle_{\mathcal{H}_{X}}$. 
%Here, uppercase symbols represent random variables, while their lowercase counterparts denote specific realizations.
We introduce $M(\x)$ as a function indicating whether a data point $(\x,y)$ is poisoned ($M(\x)=1$) or not ($M(\x)=0$). 

Given a poisoned dataset $\mathcal{D}_{bd}$, 
we consider a poisoned classifier that  minimizes the expected squared loss over this poisoned dataset:
\begin{equation} 
\label{mse}
h_{bd} = \argmin_{h\in\mathcal{H}_{X}}\mathbb{E}_{(\bm{X},Y)\sim\mathcal{D}_{bd}}(h(\bm{X})-Y)^2.
\end{equation}
Then, we provide the following Theorem: 

\begin{thm}
Assume that $\phi_{X}(\x_i)\neq \phi_{X}(\x_j)$ if $\x_i\neq \x_j$. Given a poisoned model $h_{bd}$ defined in Eq.~(\ref{mse}), there exists non-trivial linear projection operators $P_{y}$ for $y$ such that
$$\text{Cov}(\langle \phi^y_{X}(\bm{X}), h_{bd}\rangle_{\mathcal{H}_{X}}|Y=y, M(\bm{X})|Y=y)=0,$$
where $\phi^y_{X}(\x) = P_y\phi_{X}(\x)$ is the projected feature of $\phi_{X}(\x)$ for the sample pair $(\x,y)$. 
\label{theorem1}
\end{thm}

The proof of Theorem~\ref{theorem1} can be found in Appendix A. The assumption in Theorem~\ref{theorem1} requires that the mapping from the input space to the feature space preserves the uniqueness of each sample and the feature of a poisoned sample is not completely overwritten by the trigger. Then, Theorem~\ref{theorem1} states that \emph{it is possible to construct a class-conditional neural polarizer for each class that decorrelates the prediction of $h_{bd}$ from the poison status of the sample without altering the trained classifier itself.}  Such an operator can be derived from the eigenfunctions associated with the class-conditional covariance operator between $M(\bm{X})$ and $\bm{X}$ in the specific kernel space \cite{wei2023mean}.  Therefore, Theorem~\ref{theorem1} supports the existence of the class-conditional neural polarizer.

\subsubsection{Mechanism of CNPD}
Building upon the backdoor risk introduced in \cite{wei2023shared}, we demonstrate that the CNPD learning process is equivalent to minimizing an upper bound on backdoor risk, providing a theoretical guarantee of CNPD’s effectiveness.

We first recall that the \textbf{backdoor risk} $\mathcal{R}_{bd}$ defined in \cite{wei2023shared} is:
\begin{align}
    \label{eq::bd_risk}
    \mathcal{R}_{bd}(h) = \frac{\sum_{\x\sim\mathcal{D}} \mathbb{I}(h(\x_{\Delta})=T)}{|\mathcal{D}|}.
\end{align}
Here, $\mathbb{I}$ is the indicator function, which equals 1 if the argument is true and 0 otherwise,  $\mathcal{D}$ is a non-target dataset that includes samples from non-target classes, and $|\cdot|$ denotes the size of the set.

Given a backdoored model $h_{bd}$, for a poisoned sample $\x_{\Delta}$, the corresponding CNPD is trained to minimize the targeted adversarial risk for target $h_{bd}(\x_{\Delta})$. Therefore, following the work of \cite{wei2023shared, wang2019improving}, we defined the following risk for CNPD, which measures the risk of classifying an adversarial example to the target $h_{bd}(\x_{\Delta})$:

\noindent\resizebox{.85\linewidth}{!}{
  \begin{minipage}{\linewidth}
   \begin{equation*}
    \mathcal{R}_{cnpd}(h)=\frac{\sum\limits_{\x\sim\mathcal{D},t\in[1,C]}\max\limits_{\|\delta\|_{p}\leq\rho}\mathbb{I}(h_{bd}(\x_{\Delta})=t)\mathbb{I}(h(\x+\vdelta)=t,t\neq y)}{|\mathcal{D}|}.
\end{equation*} 
\end{minipage}}

\noindent We then establish the following theorem:

\begin{thm}
Assume that $h_{bd}$ is a fully backdoored model, \ie, $h_{bd}(\x_{\Delta})=T$ for all $\x\in\mathcal{D}$. Furthermore, assume that $\|\x_{\Delta}-\x\|_p\leq \rho$. Then, the following inequality holds:
\begin{equation}
    \mathcal{R}_{bd}(h)\leq \mathcal{R}_{cnpd}(h).
\end{equation}
\label{theorem2}
\end{thm}

The above theorem indicates that CNPD offers an upper bound on backdoor risk. The proof of Theorem~\ref{theorem2} can be found in Appendix A.

\section{Experiments\label{sec4}}
% \usepackage{booktabs}
\begin{table*}[t]
\centering
\caption{Defense Results on CIFAR-10 with PreAct-ResNet18 and poisoning ratio $10.0\%$.}
\label{table1}
\setlength{\tabcolsep}{3pt} % Default value: 6pt
\scalebox{0.75}{
\begin{tabular}{c|ccc|ccc|ccc|ccc|ccc|ccc|ccc}
\toprule
Defense $\rightarrow$ & \multicolumn{3}{c|}{No Defense} & \multicolumn{3}{c|}{FP \cite{liu2018fine}} & \multicolumn{3}{c|}{NAD \cite{li2021neural}} & \multicolumn{3}{c|}{NC \cite{wang2019neural}} & \multicolumn{3}{c|}{ANP \cite{wu2021adversarial}} & \multicolumn{3}{c|}{i-BAU \cite{zeng2022adversarial}} & \multicolumn{3}{c}{EP \cite{zheng2022pre}} \\ 
Attack $\downarrow$ & \multicolumn{1}{c}{ACC} & \multicolumn{1}{c}{ASR} & \multicolumn{1}{c|}{DER} & \multicolumn{1}{c}{ACC} & \multicolumn{1}{c}{ASR} & \multicolumn{1}{c|}{DER} & \multicolumn{1}{c}{ACC} & \multicolumn{1}{c}{ASR} & \multicolumn{1}{c|}{DER} & \multicolumn{1}{c}{ACC} & \multicolumn{1}{c}{ASR} & \multicolumn{1}{c|}{DER} & \multicolumn{1}{c}{ACC} & \multicolumn{1}{c}{ASR} & \multicolumn{1}{c|}{DER} & \multicolumn{1}{c}{ACC} & \multicolumn{1}{c}{ASR} & \multicolumn{1}{c|}{DER} & \multicolumn{1}{c}{ACC} & \multicolumn{1}{c}{ASR} & \multicolumn{1}{c}{DER} \\ \midrule
BadNets-A2O \cite{gu2019badnets} & $91.82$& $93.79$&  N/A & $\cellcolor[HTML]{D7E8F2}91.77$& $\cellcolor[HTML]{FEEBB8}0.84$& $\cellcolor[HTML]{F1B9B6}96.45$& $88.82$& $1.96$& $94.42$& $90.33$& $2.01$& $95.14$& $\cellcolor[HTML]{FEEBB8}92.17$& $39.14$& $77.32$& $87.43$& $4.48$& $92.46$& $89.8$& $1.26$& $95.26$\\
BadNets-A2A \cite{gu2019badnets} & $91.89$& $74.42$&  N/A & $\cellcolor[HTML]{FEEBB8}92.05$& $1.31$& $\cellcolor[HTML]{FEEBB8}86.56$& $90.73$& $1.61$& $85.82$& $89.79$& $1.11$& $85.60$& $\cellcolor[HTML]{F1B9B6}92.22$& $3.39$& $85.52$& $89.39$& $1.29$& $85.32$& $88.72$& $3.0$& $84.12$\\
Blended \cite{chen2017targeted} & $93.69$& $99.76$&  N/A & $\cellcolor[HTML]{D7E8F2}92.74$& $10.17$& $94.32$& $92.25$& $47.64$& $75.34$& $\cellcolor[HTML]{F1B9B6}93.69$& $99.76$& $50.00$& $\cellcolor[HTML]{FEEBB8}93.45$& $47.14$& $76.19$& $89.43$& $26.82$& $84.34$& $91.94$& $48.22$& $74.89$\\
Input-Aware \cite{nguyen2020input} & $94.03$& $98.35$&  N/A & $\cellcolor[HTML]{D7E8F2}94.05$& $1.62$& $98.36$& $\cellcolor[HTML]{F1B9B6}94.08$& $0.92$& $\cellcolor[HTML]{FEEBB8}98.71$& $94.00$& $0.7$& $\cellcolor[HTML]{F1B9B6}98.81$& $\cellcolor[HTML]{FEEBB8}94.06$& $1.57$& $98.39$& $89.91$& $\cellcolor[HTML]{FEEBB8}0.02$& $97.1$& $93.68$& $2.88$& $97.56$\\
LF \cite{zeng2021rethinking} & $93.01$& $99.06$&  N/A & $92.05$& $21.32$& $88.39$& $91.72$& $75.47$& $61.15$& $\cellcolor[HTML]{FEEBB8}93.01$& $99.06$& $50.00$& $\cellcolor[HTML]{F1B9B6}93.05$& $97.69$& $50.68$& $88.92$& $11.99$& $91.49$& $91.97$& $84.73$& $56.64$\\
SSBA \cite{li2021invisible} & $92.88$& $97.07$&  N/A & $\cellcolor[HTML]{D7E8F2}92.21$& $20.27$& $88.07$& $92.15$& $70.77$& $62.78$& $\cellcolor[HTML]{FEEBB8}92.88$& $97.07$& $50.00$& $\cellcolor[HTML]{F1B9B6}92.94$& $75.22$& $60.92$& $86.53$& $2.89$& $93.91$& $91.67$& $4.33$& $95.76$\\
Trojan \cite{Trojannn} & $93.47$& $99.99$&  N/A & $92.24$& $67.73$& $65.51$& $92.18$& $5.77$& $96.47$& $92.03$& $2.77$& $97.89$& $\cellcolor[HTML]{F1B9B6}93.20$& $96.12$& $51.8$& $89.29$& $\cellcolor[HTML]{D7E8F2}0.54$& $97.63$& $92.32$& $2.49$& $98.18$\\
WaNet \cite{nguyen2021wanet} & $92.8$& $98.9$&  N/A & $\cellcolor[HTML]{D7E8F2}92.94$& $\cellcolor[HTML]{F1B9B6}0.66$& $\cellcolor[HTML]{F1B9B6}99.12$& $\cellcolor[HTML]{FEEBB8}93.07$& $\cellcolor[HTML]{FEEBB8}0.73$& $\cellcolor[HTML]{FEEBB8}99.08$& $92.80$& $98.9$& $50.00$& $\cellcolor[HTML]{F1B9B6}93.76$& $1.4$& $98.75$& $90.7$& $0.88$& $97.96$& $90.47$& $96.52$& $50.03$\\
FTrojan \cite{wang2022invisible} & $93.66$& $99.99$&  N/A & $92.67$& $1.31$& $98.84$& $92.35$& $4.10$& $97.29$& $92.21$& $3.86$& $97.34$& $\cellcolor[HTML]{D7E8F2}93.16$& $0.9$& $99.29$& $87.33$& $4.58$& $94.54$& $92.47$& $1.36$& $98.72$\\
Adap-Blend \cite{qi2023revisiting} & $92.34$& $74.28$&  N/A & $\cellcolor[HTML]{D7E8F2}91.69$& $5.30$& $84.16$& $\cellcolor[HTML]{FEEBB8}92.02$& $44.87$& $64.55$& $\cellcolor[HTML]{F1B9B6}92.34$& $74.28$& $50.00$& $91.51$& $9.34$& $82.05$& $88.01$& $12.41$& $78.77$& $91.49$& $77.36$& $49.58$\\ \midrule
Average & $92.96$& $93.56$&  N/A & $\cellcolor[HTML]{FEEBB8}92.44$& $13.05$& $89.98$& $91.94$& $25.38$& $83.56$& $92.31$& $47.95$& $72.48$& $\cellcolor[HTML]{F1B9B6}92.95$& $37.19$& $78.09$& $88.69$& $6.59$& $91.35$& $91.45$& $32.21$& $80.07$\\
\toprule
\toprule
Defense $\rightarrow$ & \multicolumn{3}{c|}{FT-SAM \cite{zhu2023enhancing}} & \multicolumn{3}{c|}{SAU \cite{wei2023shared}} & \multicolumn{3}{c|}{FST \cite{min2024towards}} & \multicolumn{3}{c|}{NPD (\textbf{Ours})} & \multicolumn{3}{c|}{r-CNPD (\textbf{Ours})} & \multicolumn{3}{c|}{e-CNPD (\textbf{Ours})} & \multicolumn{3}{c}{a-CNPD (\textbf{Ours})} \\ 
Attack $\downarrow$ & \multicolumn{1}{c}{ACC} & \multicolumn{1}{c}{ASR} & \multicolumn{1}{c|}{DER} & \multicolumn{1}{c}{ACC} & \multicolumn{1}{c}{ASR} & \multicolumn{1}{c|}{DER} & \multicolumn{1}{c}{ACC} & \multicolumn{1}{c}{ASR} & \multicolumn{1}{c|}{DER} & \multicolumn{1}{c}{ACC} & \multicolumn{1}{c}{ASR} & \multicolumn{1}{c|}{DER} & \multicolumn{1}{c}{ACC} & \multicolumn{1}{c}{ASR} & \multicolumn{1}{c|}{DER} & \multicolumn{1}{c}{ACC} & \multicolumn{1}{c}{ASR} & \multicolumn{1}{c|}{DER} & \multicolumn{1}{c}{ACC} & \multicolumn{1}{c}{ASR} & \multicolumn{1}{c}{DER} \\ \midrule
BadNets-A2O \cite{gu2019badnets} & $\cellcolor[HTML]{F1B9B6}92.21$& $1.63$& $\cellcolor[HTML]{D7E8F2}96.08$& $89.81$& $1.3$& $95.24$& $91.06$& $1.66$& $95.69$& $88.93$& $1.26$& $94.82$& $89.73$& $\cellcolor[HTML]{D7E8F2}0.88$& $95.41$& $89.41$& $0.96$& $95.21$& $90.68$& $\cellcolor[HTML]{F1B9B6}0.46$& $\cellcolor[HTML]{FEEBB8}96.10$\\
BadNets-A2A \cite{gu2019badnets} & $\cellcolor[HTML]{D7E8F2}91.87$& $\cellcolor[HTML]{D7E8F2}1.03$& $\cellcolor[HTML]{F1B9B6}86.68$& $90.95$& $1.3$& $86.09$& $90.42$& $1.65$& $85.65$& $91.41$& $\cellcolor[HTML]{FEEBB8}0.89$& $\cellcolor[HTML]{D7E8F2}86.52$& $91.11$& $2.23$& $85.70$& $91.22$& $\cellcolor[HTML]{F1B9B6}0.86$& $86.44$& $91.26$& $1.64$& $86.08$\\
Blended \cite{chen2017targeted} & $92.44$& $4.91$& $96.80$& $91.75$& $5.2$& $96.31$& $92.04$& $0.80$& $98.65$& $91.18$& $\cellcolor[HTML]{D7E8F2}0.41$& $98.42$& $92.06$& $\cellcolor[HTML]{FEEBB8}0.19$& $\cellcolor[HTML]{FEEBB8}98.97$& $92.12$& $0.59$& $\cellcolor[HTML]{D7E8F2}98.80$& $92.46$& $\cellcolor[HTML]{F1B9B6}0.07$& $\cellcolor[HTML]{F1B9B6}99.23$\\
Input-Aware \cite{nguyen2020input} & $93.76$& $1.07$& $\cellcolor[HTML]{D7E8F2}98.51$& $91.93$& $10.67$& $92.79$& $93.28$& $0.92$& $98.34$& $89.57$& $\cellcolor[HTML]{D7E8F2}0.11$& $96.89$& $92.72$& $0.24$& $98.40$& $91.93$& $1.57$& $97.34$& $92.53$& $\cellcolor[HTML]{F1B9B6}0.02$& $98.41$\\
LF \cite{zeng2021rethinking} & $92.29$& $6.43$& $95.95$& $72.68$& $2.49$& $88.12$& $91.89$& $5.42$& $96.26$& $90.06$& $\cellcolor[HTML]{F1B9B6}0.21$& $97.95$& $91.46$& $\cellcolor[HTML]{FEEBB8}0.72$& $\cellcolor[HTML]{FEEBB8}98.39$& $\cellcolor[HTML]{D7E8F2}92.37$& $\cellcolor[HTML]{D7E8F2}1.18$& $\cellcolor[HTML]{F1B9B6}98.62$& $91.39$& $1.54$& $\cellcolor[HTML]{D7E8F2}97.95$\\
SSBA \cite{li2021invisible} & $92.18$& $2.06$& $97.16$& $88.96$& $2.16$& $95.50$& $92.06$& $\cellcolor[HTML]{FEEBB8}0.54$& $\cellcolor[HTML]{FEEBB8}97.85$& $90.88$& $2.77$& $96.15$& $91.98$& $2.76$& $96.71$& $91.46$& $\cellcolor[HTML]{D7E8F2}1.21$& $\cellcolor[HTML]{D7E8F2}97.22$& $92.10$& $\cellcolor[HTML]{F1B9B6}0.04$& $\cellcolor[HTML]{F1B9B6}98.12$\\
Trojan \cite{Trojannn} & $\cellcolor[HTML]{FEEBB8}92.75$& $4.12$& $97.57$& $91.64$& $1.39$& $\cellcolor[HTML]{D7E8F2}98.38$& $92.02$& $8.93$& $94.80$& $92.37$& $6.51$& $96.19$& $\cellcolor[HTML]{D7E8F2}92.53$& $\cellcolor[HTML]{FEEBB8}0.01$& $\cellcolor[HTML]{F1B9B6}99.52$& $91.92$& $1.92$& $98.26$& $91.76$& $\cellcolor[HTML]{F1B9B6}0.01$& $\cellcolor[HTML]{FEEBB8}99.13$\\
WaNet \cite{nguyen2021wanet} & $92.87$& $0.96$& $\cellcolor[HTML]{D7E8F2}98.97$& $91.94$& $0.82$& $98.61$& $92.07$& $1.26$& $98.46$& $91.57$& $\cellcolor[HTML]{D7E8F2}0.80$& $98.43$& $91.95$& $3.29$& $97.38$& $92.48$& $2.47$& $98.05$& $91.99$& $1.08$& $98.51$\\
FTrojan \cite{wang2022invisible} & $91.79$& $1.26$& $98.43$& $90.67$& $0.79$& $98.10$& $91.74$& $0.00$& $99.03$& $91.29$& $1.91$& $97.85$& $\cellcolor[HTML]{FEEBB8}93.32$& $\cellcolor[HTML]{D7E8F2}0.00$& $\cellcolor[HTML]{FEEBB8}99.82$& $92.81$& $\cellcolor[HTML]{FEEBB8}0.00$& $\cellcolor[HTML]{D7E8F2}99.57$& $\cellcolor[HTML]{F1B9B6}93.34$& $\cellcolor[HTML]{F1B9B6}0.00$& $\cellcolor[HTML]{F1B9B6}99.83$\\
Adap-Blend \cite{qi2023revisiting} & $91.33$& $7.89$& $82.69$& $86.74$& $5.14$& $81.77$& $89.62$& $\cellcolor[HTML]{D7E8F2}1.94$& $84.81$& $90.13$& $3.25$& $84.41$& $90.77$& $2.14$& $\cellcolor[HTML]{FEEBB8}85.28$& $90.08$& $\cellcolor[HTML]{FEEBB8}1.94$& $\cellcolor[HTML]{D7E8F2}85.04$& $90.30$& $\cellcolor[HTML]{F1B9B6}1.31$& $\cellcolor[HTML]{F1B9B6}85.46$\\ \midrule
Average & $\cellcolor[HTML]{D7E8F2}92.35$& $3.14$& $94.88$& $88.71$& $3.13$& $93.09$& $91.62$& $2.31$& $94.95$& $90.74$& $1.81$& $94.76$& $91.76$& $\cellcolor[HTML]{FEEBB8}1.25$& $\cellcolor[HTML]{FEEBB8}95.56$& $91.58$& $\cellcolor[HTML]{D7E8F2}1.27$& $\cellcolor[HTML]{D7E8F2}95.45$& $91.78$& $\cellcolor[HTML]{F1B9B6}0.62$& $\cellcolor[HTML]{F1B9B6}95.88$\\
\bottomrule
\end{tabular}}
\end{table*}

To evaluate the effectiveness of the proposed method, we perform extensive experiments across different benchmark datasets and network architectures, including CIFAR-10 \cite{krizhevsky2009learning}, Tiny ImageNet \cite{le2015tiny}, and GTSRB \cite{stallkamp2011german} datasets, on PreAct-ResNet18 \cite{he2016identity} and VGG19-BN \cite{simonyan2014very} networks. We evaluate the proposed method against ten state-of-the-art (SOTA) backdoor attack methods, and compare the performance with nine SOTA backdoor defense methods. Additionally, we provide some analyses in Sec. \ref{sec4.4}.

\subsection{Implementation Details}
\subsubsection{Datasets and models}
Three benchmark datasets \cite{wubackdoorbench} are used to conduct comparison experiments including CIFAR-10 \cite{krizhevsky2009learning}, Tiny ImageNet \cite{le2015tiny}, and GTSRB \cite{stallkamp2011german}. In details, 
CIFAR-10 contains a total of 60,000 images divided into ten categories. Each category has 5,000 and 1000 images for training and testing, respectively. The size of each image is $32 \times 32$.
Tiny ImageNet, which is a smaller version of the ImageNet dataset \cite{deng2009imagenet}, includes 100,000 images for training and an additional 10,000 for testing, spread across 200 different classes. The images in this dataset have $64 \times 64$ pixels in size.
GTSRB contains a total of 39,209 images for training and 12,630 for testing, grouped into 43 categories. Each image in GTSRB is $32 \times 32$ pixels in dimensions.
To demonstrate the applicability of our method across different network architectures, we conduct experiments on both PreAct-ResNet18 \cite{he2016identity} and VGG19-BN \cite{simonyan2014very}, and compare their performance with SOTA methods.

\subsubsection{Attack settings}
We evaluate the proposed three defense methods against ten well-known backdoor attacks, including BadNets \cite{gu2019badnets} (both BadNets-A2O and BadNets-A2A, representing attacking one target label and multiple target labels, respectively), Blended attack (Blended \cite{chen2017targeted}), Input-aware dynamic backdoor attack (Input-aware \cite{nguyen2020input}), Low frequency attack (LF \cite{zeng2021rethinking}), Sample-specific backdoor attack (SSBA \cite{li2021invisible}), Trojan backdoor attack (Trojannn \cite{Trojannn}), Warping-based poisoned networks (WaNet \cite{nguyen2021wanet}), Frequency domain trojan attack (FTrojan \cite{wang2022invisible}), and Adaptive Backdoor Attack (Adap-Blend \cite{qi2023revisiting}). To ensure a fair comparison, we follow  the default attack configurations outlined in BackdoorBench \cite{wubackdoorbench}. The poisoning ratio is set to 10\% and the targeted label is set to the $0^{th}$ label when evaluating against state-of-the-art defenses. To demonstrate the effectiveness of our method under multiple target labels setting, we employ an all-to-all label attack strategy, BadNets-A2A, where the target labels for original labels $y$ are set to $y_t=(y+1) \bmod C$. Detailed introductions about these attacks are provided in Appendix B.

\subsubsection{Defense settings}
We compare the proposed three defense methods against nine SOTA backdoor defense methods, \ie, FP \cite{liu2018fine}, NAD \cite{li2021neural}, NC \cite{wang2019neural}, ANP \cite{wu2021adversarial}, i-BAU \cite{zeng2022adversarial}, EP \cite{zheng2022pre}, FT-SAM \cite{zhu2023enhancing}, SAU \cite{wei2023shared}, and FST \cite{min2024towards}. All these defenses have access to 5\% benign training samples for training. All the training hyperparameters are aligned with BackdoorBench \cite{wubackdoorbench}. We evaluate the proposed three defense methods with these SOTA defenses on different datasets and networks. 

\subsubsection{Details of our methods}
For our methods, we apply an $l_2$ norm constraint to the adversarial perturbations, with a perturbation bound of 3 for the CIFAR-10 and GTSRB datasets, and 6 for the Tiny ImageNet dataset. The loss hyperparameters $\lambda_1$, $\lambda_2$, and $\lambda_3$ are set to $1.0$, $0.4$, and $0.4$ for the CIFAR-10 in our NPD, e-CNPD and a-CNPD, and set to $1.0$, $0.5$, and $0.5$ for in our r-CNPD. We train the neural polarizer for 50, 10, 100, and 200 epochs for the NPD, a-CNPD, e-CNPD, and r-CNPD methods, respectively, with a learning rate of $0.01$ on the CIFAR-10 dataset. The neural polarizer block is inserted before the final convolutional layer of the PreAct-ResNet18 architecture for CNPD and before the penultimate convolutional layer for NPD. Additional implementation details regarding state-of-the-art attacks, defenses, and our methods can be found in Appendix B.

\subsubsection{Evaluation metrics}
In this work, we utilize three primary evaluation metrics to assess the effectiveness of various defense methods: clean Accuracy (ACC), Attack Success Rate (ASR), and Defense Effectiveness Rating (DER). 
\begin{itemize}
    \item \textbf{ACC} reflects the accuracy of benign samples.
    \item \textbf{ASR} evaluates the percentage of poisoned samples that are successfully classified as the target label.
    \item \textbf{DER} \cite{zhu2023enhancing}) is a comprehensive metric that considers both ACC and ASR:
\end{itemize}
A higher ACC, a lower ASR, and a higher DER indicate superior defense performance. 
In the main tables, the effectiveness of defense methods against various attack methods is color-coded for clarity: the most effective defense is highlighted in $\cellcolor[HTML]{F1B9B6}\text{red}$, the second most effective in $\cellcolor[HTML]{FEEBB8}\text{yellow}$, and the third in $\cellcolor[HTML]{D7E8F2}\text{blue}$.

% \usepackage{booktabs}
\begin{table*}[t]
\centering
\caption{Defense Results on GTSRB with PreAct-ResNet18 and poisoning ratio $10.0\%$.}
\label{table2}
\setlength{\tabcolsep}{3pt} % Default value: 6pt
\scalebox{0.75}{
\begin{tabular}{c|ccc|ccc|ccc|ccc|ccc|ccc|ccc}
\toprule
Defense $\rightarrow$ & \multicolumn{3}{c|}{No Defense} & \multicolumn{3}{c|}{FP \cite{liu2018fine}} & \multicolumn{3}{c|}{NAD \cite{li2021neural}} & \multicolumn{3}{c|}{NC \cite{wang2019neural}} & \multicolumn{3}{c|}{ANP \cite{wu2021adversarial}} & \multicolumn{3}{c|}{i-BAU \cite{zeng2022adversarial}} & \multicolumn{3}{c}{EP \cite{zheng2022pre}} \\ 
Attack $\downarrow$ & \multicolumn{1}{c}{ACC} & \multicolumn{1}{c}{ASR} & \multicolumn{1}{c|}{DER} & \multicolumn{1}{c}{ACC} & \multicolumn{1}{c}{ASR} & \multicolumn{1}{c|}{DER} & \multicolumn{1}{c}{ACC} & \multicolumn{1}{c}{ASR} & \multicolumn{1}{c|}{DER} & \multicolumn{1}{c}{ACC} & \multicolumn{1}{c}{ASR} & \multicolumn{1}{c|}{DER} & \multicolumn{1}{c}{ACC} & \multicolumn{1}{c}{ASR} & \multicolumn{1}{c|}{DER} & \multicolumn{1}{c}{ACC} & \multicolumn{1}{c}{ASR} & \multicolumn{1}{c|}{DER} & \multicolumn{1}{c}{ACC} & \multicolumn{1}{c}{ASR} & \multicolumn{1}{c}{DER} \\ \midrule
BadNets-A2O \cite{gu2019badnets} & $96.35$& $95.02$&  N/A & $\cellcolor[HTML]{F1B9B6}98.12$& $0.0$& $\cellcolor[HTML]{D7E8F2}97.51$& $97.54$& $79.94$& $57.54$& $93.21$& $0.02$& $95.93$& $96.52$& $18.49$& $88.27$& $96.35$& $0.00$& $\cellcolor[HTML]{FEEBB8}97.51$& $96.53$& $1.38$& $96.82$\\
BadNets-A2A \cite{gu2019badnets} & $97.05$& $92.33$&  N/A & $\cellcolor[HTML]{D7E8F2}98.11$& $0.51$& $\cellcolor[HTML]{D7E8F2}95.91$& $97.84$& $2.46$& $94.93$& $94.05$& $0.5$& $94.41$& $97.07$& $92.01$& $50.16$& $95.3$& $0.43$& $95.08$& $96.45$& $1.4$& $95.17$\\
Blended \cite{chen2017targeted} & $98.17$& $100.0$&  N/A & $\cellcolor[HTML]{FEEBB8}98.20$& $68.45$& $65.78$& $\cellcolor[HTML]{D7E8F2}97.98$& $99.29$& $50.26$& $77.48$& $8.77$& $85.27$& $\cellcolor[HTML]{F1B9B6}98.25$& $99.94$& $50.03$& $95.04$& $96.39$& $50.24$& $95.42$& $100.0$& $48.63$\\
Input-Aware \cite{nguyen2020input} & $97.17$& $97.09$&  N/A & $\cellcolor[HTML]{FEEBB8}97.98$& $0.22$& $\cellcolor[HTML]{D7E8F2}98.43$& $\cellcolor[HTML]{D7E8F2}97.47$& $65.55$& $65.77$& $95.42$& $0.03$& $97.65$& $96.20$& $\cellcolor[HTML]{D7E8F2}0.00$& $98.06$& $96.03$& $\cellcolor[HTML]{FEEBB8}0.00$& $97.98$& $92.98$& $0.1$& $96.4$\\
LF \cite{zeng2021rethinking} & $97.97$& $99.58$&  N/A & $\cellcolor[HTML]{D7E8F2}97.87$& $69.19$& $65.15$& $\cellcolor[HTML]{F1B9B6}98.24$& $79.76$& $59.91$& $90.73$& $0.05$& $96.15$& $\cellcolor[HTML]{FEEBB8}98.19$& $89.00$& $55.29$& $88.69$& $7.43$& $91.44$& $96.4$& $99.15$& $49.43$\\
SSBA \cite{li2021invisible} & $98.31$& $99.77$&  N/A & $\cellcolor[HTML]{F1B9B6}98.47$& $60.19$& $69.79$& $\cellcolor[HTML]{FEEBB8}98.37$& $96.95$& $51.41$& $89.45$& $2.43$& $94.24$& $\cellcolor[HTML]{D7E8F2}98.35$& $99.71$& $50.03$& $87.27$& $\cellcolor[HTML]{D7E8F2}0.18$& $94.27$& $97.59$& $99.32$& $49.87$\\
Trojan \cite{Trojannn} & $98.33$& $100.0$&  N/A & $98.00$& $42.08$& $78.80$& $\cellcolor[HTML]{FEEBB8}98.01$& $0.10$& $\cellcolor[HTML]{D7E8F2}99.79$& $90.39$& $0.36$& $95.85$& $\cellcolor[HTML]{F1B9B6}98.31$& $99.91$& $50.03$& $93.66$& $0.00$& $97.66$& $97.46$& $0.05$& $99.54$\\
WaNet \cite{nguyen2021wanet} & $95.71$& $98.2$&  N/A & $\cellcolor[HTML]{FEEBB8}98.88$& $0.28$& $98.96$& $98.32$& $\cellcolor[HTML]{D7E8F2}0.04$& $\cellcolor[HTML]{D7E8F2}99.08$& $96.57$& $0.15$& $99.02$& $95.54$& $4.34$& $96.85$& $97.5$& $0.26$& $98.97$& $97.26$& $26.92$& $85.64$\\
FTrojan \cite{wang2022invisible} & $98.54$& $100.0$&  N/A & $\cellcolor[HTML]{D7E8F2}98.46$& $99.99$& $49.96$& $97.98$& $0.01$& $99.71$& $96.22$& $100.0$& $48.84$& $98.43$& $0.00$& $\cellcolor[HTML]{D7E8F2}99.94$& $94.28$& $0.32$& $97.71$& $98.05$& $0.0$& $99.75$\\
Adap-Blend \cite{qi2023revisiting} & $97.28$& $79.31$&  N/A & $\cellcolor[HTML]{FEEBB8}98.50$& $0.42$& $89.44$& $98.00$& $45.76$& $66.77$& $90.29$& $93.29$& $46.51$& $97.05$& $64.08$& $57.50$& $96.12$& $10.05$& $84.05$& $95.71$& $84.6$& $49.22$\\ \midrule
Average & $97.49$& $96.13$&  N/A & $\cellcolor[HTML]{F1B9B6}98.26$& $34.13$& $80.97$& $\cellcolor[HTML]{FEEBB8}97.98$& $46.99$& $74.52$& $91.38$& $20.56$& $85.39$& $97.39$& $56.75$& $69.62$& $94.02$& $11.51$& $90.49$& $96.39$& $41.29$& $77.05$\\

\toprule

\toprule
Defense $\rightarrow$ & \multicolumn{3}{c|}{FT-SAM \cite{zhu2023enhancing}} & \multicolumn{3}{c|}{FST \cite{min2024towards}} & \multicolumn{3}{c|}{SAU \cite{wei2023shared}} & \multicolumn{3}{c|}{NPD (\textbf{Ours})} & \multicolumn{3}{c|}{r-CNPD (\textbf{Ours})} & \multicolumn{3}{c|}{e-CNPD (\textbf{Ours})} & \multicolumn{3}{c}{a-CNPD (\textbf{Ours})} \\ 
Attack $\downarrow$ & \multicolumn{1}{c}{ACC} & \multicolumn{1}{c}{ASR} & \multicolumn{1}{c|}{DER} & \multicolumn{1}{c}{ACC} & \multicolumn{1}{c}{ASR} & \multicolumn{1}{c|}{DER} & \multicolumn{1}{c}{ACC} & \multicolumn{1}{c}{ASR} & \multicolumn{1}{c|}{DER} & \multicolumn{1}{c}{ACC} & \multicolumn{1}{c}{ASR} & \multicolumn{1}{c|}{DER} & \multicolumn{1}{c}{ACC} & \multicolumn{1}{c}{ASR} & \multicolumn{1}{c|}{DER} & \multicolumn{1}{c}{ACC} & \multicolumn{1}{c}{ASR} & \multicolumn{1}{c|}{DER} & \multicolumn{1}{c}{ACC} & \multicolumn{1}{c}{ASR} & \multicolumn{1}{c}{DER} \\ \midrule
BadNets-A2O \cite{gu2019badnets} & $96.36$& $0.17$& $97.43$& $\cellcolor[HTML]{FEEBB8}98.02$& $0.02$& $97.50$& $96.03$& $\cellcolor[HTML]{D7E8F2}0.00$& $97.35$& $95.89$& $4.62$& $94.97$& $\cellcolor[HTML]{D7E8F2}97.64$& $0.45$& $97.29$& $96.04$& $\cellcolor[HTML]{FEEBB8}0.00$& $97.36$& $97.08$& $\cellcolor[HTML]{F1B9B6}0.00$& $\cellcolor[HTML]{F1B9B6}97.51$\\
BadNets-A2A \cite{gu2019badnets} & $\cellcolor[HTML]{F1B9B6}98.27$& $\cellcolor[HTML]{FEEBB8}0.30$& $\cellcolor[HTML]{FEEBB8}96.01$& $\cellcolor[HTML]{FEEBB8}98.12$& $\cellcolor[HTML]{F1B9B6}0.26$& $\cellcolor[HTML]{F1B9B6}96.03$& $95.84$& $\cellcolor[HTML]{D7E8F2}0.31$& $95.40$& $97.09$& $9.33$& $91.50$& $97.34$& $2.43$& $94.95$& $97.24$& $1.09$& $95.62$& $97.79$& $0.53$& $95.90$\\
Blended \cite{chen2017targeted} & $95.84$& $11.50$& $93.08$& $97.89$& $77.26$& $61.23$& $96.29$& $30.00$& $84.06$& $97.32$& $2.44$& $98.35$& $97.53$& $\cellcolor[HTML]{D7E8F2}0.52$& $\cellcolor[HTML]{D7E8F2}99.42$& $97.15$& $\cellcolor[HTML]{FEEBB8}0.00$& $\cellcolor[HTML]{FEEBB8}99.49$& $97.62$& $\cellcolor[HTML]{F1B9B6}0.00$& $\cellcolor[HTML]{F1B9B6}99.73$\\
Input-Aware \cite{nguyen2020input} & $\cellcolor[HTML]{F1B9B6}98.23$& $0.02$& $\cellcolor[HTML]{F1B9B6}98.54$& $97.01$& $\cellcolor[HTML]{F1B9B6}0.00$& $\cellcolor[HTML]{FEEBB8}98.47$& $97.27$& $1.14$& $97.97$& $95.72$& $1.72$& $96.96$& $94.35$& $1.88$& $96.20$& $97.22$& $1.03$& $98.03$& $97.25$& $0.97$& $98.06$\\
LF \cite{zeng2021rethinking} & $97.76$& $2.55$& $98.41$& $97.01$& $1.82$& $98.40$& $96.37$& $\cellcolor[HTML]{D7E8F2}0.04$& $98.97$& $97.3$& $0.77$& $\cellcolor[HTML]{D7E8F2}99.07$& $95.41$& $1.12$& $97.95$& $97.72$& $\cellcolor[HTML]{FEEBB8}0.00$& $\cellcolor[HTML]{F1B9B6}99.67$& $97.50$& $\cellcolor[HTML]{F1B9B6}0.00$& $\cellcolor[HTML]{FEEBB8}99.56$\\
SSBA \cite{li2021invisible} & $95.99$& $0.70$& $98.37$& $97.75$& $32.55$& $83.33$& $95.38$& $1.95$& $97.45$& $98.02$& $3.45$& $98.02$& $98.15$& $0.29$& $\cellcolor[HTML]{FEEBB8}99.66$& $97.8$& $\cellcolor[HTML]{F1B9B6}0.00$& $\cellcolor[HTML]{D7E8F2}99.63$& $98.12$& $\cellcolor[HTML]{FEEBB8}0.02$& $\cellcolor[HTML]{F1B9B6}99.78$\\
Trojan \cite{Trojannn} & $96.92$& $0.11$& $99.24$& $97.88$& $3.55$& $98.00$& $96.01$& $0.07$& $98.80$& $96.47$& $\cellcolor[HTML]{D7E8F2}0.00$& $99.07$& $97.86$& $0.54$& $99.50$& $97.92$& $\cellcolor[HTML]{FEEBB8}0.00$& $\cellcolor[HTML]{FEEBB8}99.79$& $\cellcolor[HTML]{D7E8F2}98.00$& $\cellcolor[HTML]{F1B9B6}0.00$& $\cellcolor[HTML]{F1B9B6}99.83$\\
WaNet \cite{nguyen2021wanet} & $\cellcolor[HTML]{D7E8F2}98.61$& $\cellcolor[HTML]{FEEBB8}0.00$& $\cellcolor[HTML]{FEEBB8}99.10$& $\cellcolor[HTML]{F1B9B6}98.88$& $\cellcolor[HTML]{F1B9B6}0.00$& $\cellcolor[HTML]{F1B9B6}99.10$& $97.55$& $0.08$& $99.06$& $97.74$& $7.77$& $95.21$& $96.81$& $1.37$& $98.42$& $97.9$& $0.46$& $98.87$& $97.70$& $0.15$& $99.02$\\
FTrojan \cite{wang2022invisible} & $\cellcolor[HTML]{F1B9B6}98.62$& $0.00$& $\cellcolor[HTML]{F1B9B6}100.00$& $\cellcolor[HTML]{FEEBB8}98.54$& $0.01$& $\cellcolor[HTML]{FEEBB8}99.99$& $97.81$& $\cellcolor[HTML]{D7E8F2}0.00$& $99.63$& $96.3$& $6.24$& $95.76$& $98.30$& $0.41$& $99.68$& $98.19$& $\cellcolor[HTML]{FEEBB8}0.00$& $99.82$& $98.39$& $\cellcolor[HTML]{F1B9B6}0.00$& $99.92$\\
Adap-Blend \cite{qi2023revisiting} & $\cellcolor[HTML]{F1B9B6}98.69$& $9.81$& $84.75$& $98.29$& $\cellcolor[HTML]{D7E8F2}0.00$& $\cellcolor[HTML]{FEEBB8}89.65$& $97.95$& $0.10$& $\cellcolor[HTML]{D7E8F2}89.61$& $97.29$& $2.34$& $88.48$& $\cellcolor[HTML]{D7E8F2}98.29$& $4.52$& $87.39$& $97.08$& $\cellcolor[HTML]{FEEBB8}0.00$& $89.55$& $97.78$& $\cellcolor[HTML]{F1B9B6}0.00$& $\cellcolor[HTML]{F1B9B6}89.65$\\ \midrule
Average & $97.53$& $2.52$& $96.49$& $\cellcolor[HTML]{D7E8F2}97.94$& $11.55$& $92.17$& $96.65$& $3.37$& $95.83$& $96.91$& $3.87$& $95.74$& $97.17$& $\cellcolor[HTML]{D7E8F2}1.35$& $\cellcolor[HTML]{D7E8F2}97.05$& $97.42$& $\cellcolor[HTML]{FEEBB8}0.26$& $\cellcolor[HTML]{FEEBB8}97.78$& $97.72$& $\cellcolor[HTML]{F1B9B6}0.17$& $\cellcolor[HTML]{F1B9B6}97.90$\\

\bottomrule
\end{tabular}}
\end{table*}

\begin{table}[h]
\centering
\caption{Defense Results with different poisoning ratios on CIFAR-10 with PreAct-ResNet18.}
\label{table6}
\setlength{\tabcolsep}{3pt} % Default value: 6pt
\scalebox{0.75}{
\begin{tabular}{c|cc|cc|cc|cc}
\toprule
Poisoning ratio $\rightarrow$ & \multicolumn{2}{c|}{0.05} & \multicolumn{2}{c|}{0.01} & \multicolumn{2}{c|}{0.005} & \multicolumn{2}{c}{0.001} \\
Attack $\downarrow$                & ACC   & ASR  & ACC   & ASR  & ACC   & ASR  & ACC   & ASR  \\ \midrule
BadNets-A2O \cite{gu2019badnets}   & 90.6  & 0.96 & 91.82 & 0.33 & 92.75 & 1.59 & 91.95 & 1.32 \\
Blended \cite{chen2017targeted}    & 91.21 & 2.29 & 91.34 & 5.58 & 91.81 & 9.04 & 92.63 & 7.29 \\
Input-Aware \cite{nguyen2020input} & 86.03 & 3.61 & 91.07 & 6.46 & 89.77 & 1.93 & 89.95 & 1.88 \\
LF \cite{zeng2021rethinking}       & 91.37 & 5.29 & 91.73 & 3.02 & 91.43 & 8.2  & 91.5  & 5.88 \\
SSBA \cite{li2021invisible}        & 91.5  & 0.27 & 91.92 & 2.34 & 91.29 & 1.08 & 91.44 & 0.39 \\
Trojan \cite{Trojannn}             & 92.1  & 6.12 & 90.43 & 8.28 & 91.3  & 2.87 & 91.12 & 9.29 \\
WaNet \cite{nguyen2021wanet}       & 92.1  & 4.18 & 90.81 & 2.64 & 90.96 & 1.17 & 90.24 & 0.48 \\ \midrule
Average                            & 90.70 & 3.25 & 91.30 & 4.09 & 91.33 & 3.70 & 91.26 & 3.79 \\ \bottomrule
\end{tabular}}
\end{table}
\begin{table}[h]
\centering
\caption{Defense Results of e-CNPD with different clean ratios on CIFAR-10 with PreAct-ResNet18.}
\label{table7_2}
\setlength{\tabcolsep}{3pt} % Default value: 6pt
\scalebox{0.75}{
\begin{tabular}{c|cc|cc|cc|cc}
\toprule
Clean ratio $\rightarrow$ & \multicolumn{2}{c|}{0.05} & \multicolumn{2}{c|}{0.01} & \multicolumn{2}{c|}{0.005} & \multicolumn{2}{c}{0.001} \\
Attack $\downarrow$                & ACC   & ASR  & ACC   & ASR  & ACC   & ASR  & ACC   & ASR  \\ \midrule
BadNets-A2O \cite{gu2019badnets}   & 89.41 & 0.96 & 89.93 & 0.54 & 89.27 & 1.63 & 86.07 & 0.49 \\
Blended \cite{chen2017targeted}    & 92.12 & 0.59 & 91.02 & 0.01 & 91.91 & 1.05 & 86.41 & 0.24 \\
Input-Aware \cite{nguyen2020input} & 91.93 & 1.58 & 89.98 & 0.26 & 89.77 & 0.03 & 85.75 & 0.28 \\
LF \cite{zeng2021rethinking}       & 92.37 & 1.18 & 90.26 & 0.10 & 90.07 & 0.68 & 85.99 & 0.16 \\
SSBA \cite{li2021invisible}        & 91.46 & 1.21 & 90.70 & 0.38 & 90.13 & 6.54 & 87.95 & 1.62 \\
Trojan \cite{Trojannn}             & 91.92 & 1.92 & 91.09 & 0.00 & 90.91 & 3.23 & 86.93 & 0.22 \\
WaNet \cite{nguyen2021wanet}       & 92.48 & 2.47 & 91.73 & 2.01 & 91.69 & 3.35 & 83.37 & 5.09 \\ \midrule
Average                            & 91.67 & 1.42 & 90.67 & 0.47 & 90.54 & 2.36 & 86.07 & 1.16 \\ \bottomrule
\end{tabular}}
\end{table}
\begin{table}[h]
\centering
\caption{Defense Results of a-CNPD with different clean ratios on CIFAR-10 with PreAct-ResNet18.}
\label{table7_1}
\setlength{\tabcolsep}{3pt} % Default value: 6pt
\scalebox{0.75}{
\begin{tabular}{c|cc|cc|cc|cc}
\toprule
Clean ratio $\rightarrow$ & \multicolumn{2}{c|}{0.05} & \multicolumn{2}{c|}{0.01} & \multicolumn{2}{c|}{0.005} & \multicolumn{2}{c}{0.001} \\
Attack $\downarrow$                & ACC   & ASR  & ACC   & ASR   & ACC   & ASR   & ACC   & ASR   \\ \midrule
BadNets-A2O \cite{gu2019badnets}   & 90.68 & 0.46 & 90.67 & 0.56  & 90.47 & 0.50  & 88.01 & 0.06  \\
Blended \cite{chen2017targeted}    & 92.46 & 0.07 & 92.53 & 5.82  & 92.23 & 9.64  & 90.05 & 6.41  \\
Input-Aware \cite{nguyen2020input} & 92.53 & 0.02 & 90.29 & 0.32  & 90.18 & 0.29  & 88.57 & 0.81  \\
LF \cite{zeng2021rethinking}       & 91.39 & 1.54 & 91.55 & 4.36 & 91.45 & 7.44 & 88.90 & 8.10 \\
SSBA \cite{li2021invisible}        & 92.10 & 0.04 & 91.53 & 0.01  & 91.54 & 0.07  & 87.33 & 0.06  \\
Trojan \cite{Trojannn}             & 91.76 & 0.01 & 92.70 & 9.37  & 92.47 & 20.70 & 90.60 & 2.02  \\
WaNet \cite{nguyen2021wanet}       & 91.99 & 1.08 & 90.85 & 0.17  & 90.69 & 0.19  & 88.69 & 6.77  \\ \midrule
Average                            & 91.84 & 0.46 & 91.45 & 2.95 & 91.29 & 5.55 & 88.88 & 3.46  \\ \bottomrule
\end{tabular}}
\end{table}

\subsection{Main Results}
We first evaluate the effectiveness of our NPD and CNPD defense methods against ten SOTA backdoor attacks, using three different datasets and two network architectures. As shown in Tab. \ref{table1} and \ref{table2},  we observe that our defense methods demonstrate superior performance across all tested attacks. Specifically, our r-CNPD, e-CNPD, and a-CNPD methods achieved the first, second, and third highest average DER scores, respectively, and our a-CNPD achieves average DERs of 95.88\%, 97.90\%, and 93.32\% on CIFAR-10, GTSRB, and Tiny ImageNet datasets, respectively. Additionaly, our NPD also shows superior performance in terms of DER for almost all attacks compared to SOTA defenses. These results highlight that our methods outperform other defenses, with the a-CNPD method yielding the best overall average DER. Defense results with Tiny ImageNet dataset and VGG19-BN network are provided in Appendix C.

\textbf{Effectiveness Against SOTA defenses.}
Tab. \ref{table1} and \ref{table2} present a comparison between our method and the SOTA backdoor defenses. It is evident that our method consistently performs well across different datasets and models, achieving both high ACC and low ASR. While FP and ANP show high ACC, their effectiveness in reducing ASR is highly unstable, highlighting the limitations of existing pruning-based methods. In terms of reducing ASR, SAU and FST demonstrate strong performance, but they sometimes fall short in maintaining ACC. Although FT-SAM also demonstrates some advancements, it struggles with complex attacks on large datasets, \ie, Tiny ImageNet. In contrast, our method excels in both maintaining high ACC and reducing ASR, as reflected by the high DER. 

\textbf{Comparison of our defenses.}
The four defense methods proposed in this work achieve the best DER and ASR. Among them, a-CNPD generally delivers the best overall performance, while e-CNPD and r-CNPD show comparable results, ranking second and third, respectively. Specifically, a-CNPD is able to reduce the ASR to below 1\% while maintaining excellent model utility.

\subsection{Ablation Study}
\subsubsection{Effectiveness against low poisoning ratio}
Low poisoning-ratio attacks have long posed a significant challenge to the effectiveness of defense mechanisms \cite{zhang2024reliable,qi2023revisiting,ma2023dihba}. Tab. \ref{table6} presents the results of our e-CNPD defense against attacks with varying poisoning ratios on CIFAR-10 dataset. As shown, our method remains robust even under low poisoning ratios, likely because it effectively identifies backdoor-related signals and blocks them efficiently.
\begin{table}[h]
\centering
\caption{Defense Results of e-CNPD with different clean ratios on CIFAR-10 with PreAct-ResNet18.}
\label{table8}
\setlength{\tabcolsep}{3pt} % Default value: 6pt
\scalebox{0.85}{
\begin{tabular}{cccc|cc|cc|cc}
\toprule
\multicolumn{4}{c|}{Attack $\rightarrow$} &
  \multicolumn{2}{c|}{Blended \cite{chen2017targeted}} &
  \multicolumn{2}{c|}{LF \cite{zeng2021rethinking}} &
  \multicolumn{2}{c}{Trojan \cite{Trojannn}} \\
$\gL_{bn}$ & $\gL_{bce1}$ & $\gL_{bce2}$ & $\gL_{asr}$ & ACC   & ASR   & ACC   & ASR   & ACC   & ASR   \\ \hline
\checkmark &              &               &             & 93.53 & 67.24 & 92.98 & 92.43 & 93.39 & 99.97 \\
\checkmark & \checkmark   &               &             & 92.64 & 1.11  & 92.2  & 32.89 & 93.21 & 5.61  \\
\checkmark &              & \checkmark    &             & 93.13 & 15.17 & 92.32 & 10.64 & 93.13 & 6.81  \\
\checkmark &              &               & \checkmark  & 92.91 & 8.52  & 92.48 & 2.06  & 92.74 & 1.25  \\
\checkmark & \checkmark   & \checkmark    &             & 92.43 & 0.19  & 91.5  & 5.29  & 92.99 & 1.44  \\
\checkmark & \checkmark   & \checkmark    & \checkmark  & 92.46 & 0.07  & 91.39 & 1.54  & 91.76 & 0.01  \\ \bottomrule
\end{tabular}}
\end{table}
\subsubsection{Effectiveness on low clean ratio}
In practice, training samples for fine-tuning may be scarce. Tab. \ref{table7_2} and \ref{table7_1} show the defense results of our e-CNPD and a-CNPD methods under different clean ratios. As the tables indicate, our methods remain effective even with low clean ratios, although there is a slight decrease in model performance, suggesting that some regularization may be needed to maintain performance. On the other hand, the consistently low ASR demonstrates that our methods are lightweight and efficient, as they require very few training parameters, allowing for strong defense performance even with limited training samples.

\begin{figure}[h]
\centering
\includegraphics[width=0.8\linewidth]{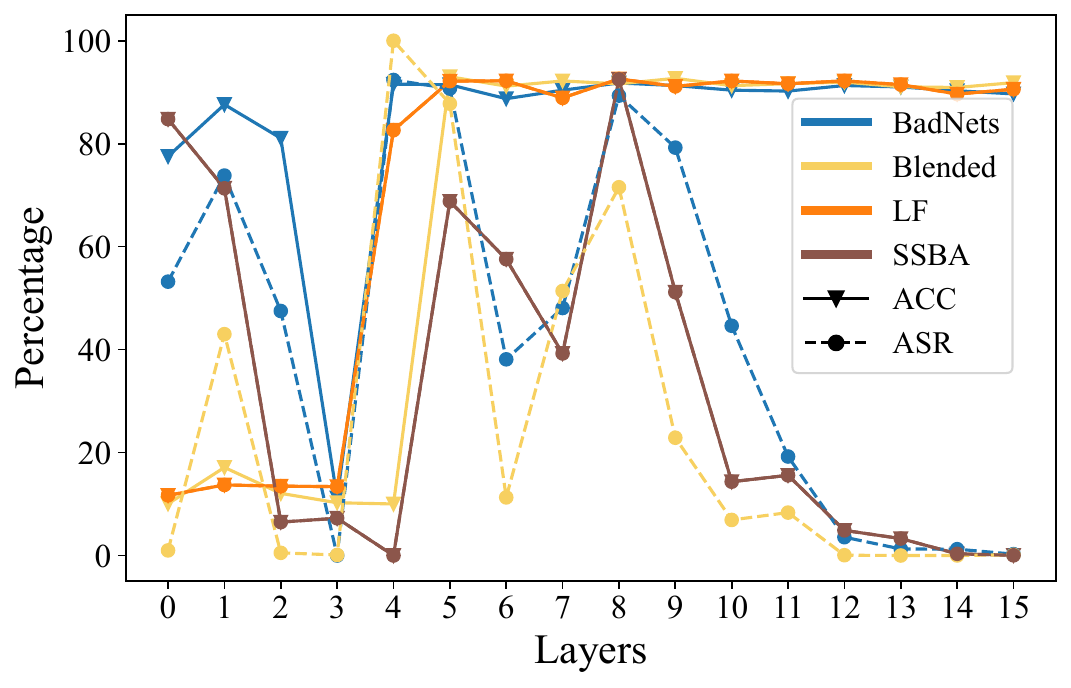}
\caption{Defense performance of inserting NP into different layers.}
\label{fig:layers}
\end{figure}

\subsubsection{Performance of inserting NP into different layers}
We evaluate the impact of selecting different layers for inserting the NP layer by positioning it before each convolutional layer in the PreAct-ResNet18 network, tested on the CIFAR-10 dataset. Fig. \ref{fig:layers} presents our a-CNPD's defense performance under different attacks. The results indicate that inserting the NP layer into shallower layers leads to a drop in accuracy, as even minor perturbations in these layers can cause significant instability in the final output. However, as the insertion point moves deeper into the network, the features become more distinct, leading to improved defense performance.

\subsubsection{Effectiveness of each loss term}
We conduct an ablation study to assess the contribution of each component of the loss function to the overall performance on CIFAR-10 dataset. Specifically, we examine the first and second terms of the loss function, $\gL_{bce}$ (see Eq. (\ref{eq:bce_loss})), which we denote as $\gL_{bce1}$ and $\gL_{bce2}$, respectively. As shown in Tab. \ref{table8}, both $\gL_{bce}$ and $\gL_{asr}$ play a crucial role in enhancing overall performance, and removing any component results in a considerable drop in defense effectiveness.

\subsection{Analysis\label{sec4.4}}
\begin{table}[h]
\centering
\caption{Detection Results on CIFAR-10 with PreAct-ResNet18.}
\label{table5}
\setlength{\tabcolsep}{2.5pt} % Default value: 6pt
\scalebox{0.65}{
\begin{tabular}{c|cc|cc|cc|cc|cc}
\toprule
Detection $\rightarrow$ &
  \multicolumn{2}{c|}{SCALE-UP \cite{guoscale}} &
  \multicolumn{2}{c|}{SentiNet \cite{chou2020sentinet}} &
  \multicolumn{2}{c|}{STRIP \cite{gao2019strip}} &
  \multicolumn{2}{c|}{IBD-PSC \cite{hou2024ibdpsc}} &
  \multicolumn{2}{c}{NPDT (\textbf{Ours})} \\
Attack $\downarrow$                & TPR   & FPR   & TPR   & FPR           & TPR   & FPR   & TPR            & FPR           & TPR            & FPR           \\ \midrule
BadNets-A2O \cite{gu2019badnets}   & 39.55 & 38.96 & 35.33 & 13.10         & 88.34 & 10.10 & 89.64          & 8.46          & \textbf{99.79} & \textbf{6.77} \\
BadNets-A2A \cite{gu2019badnets}   & 16.95 & 34.65 & 25.99 & 23.30         & 0.56  & 8.80  & 11.79          & \textbf{5.41} & \textbf{99.66} & 6.31          \\
Blended \cite{chen2017targeted}    & 60.77 & 44.31 & 2.92  & 4.20          & 59.46 & 10.00 & 99.10          & 9.19          & \textbf{99.64} & \textbf{4.16} \\
Input-Aware \cite{nguyen2020input} & 61.06 & 39.62 & 39.91 & 38.10         & 0.85  & 8.30  & 46.92          & 12.46         & \textbf{99.91} & \textbf{5.85} \\
LF \cite{zeng2021rethinking}       & 86.63 & 37.13 & 0.02  & \textbf{0.00} & 84.38 & 9.50  & 99.19          & 9.24          & \textbf{99.41} & 5.22          \\
SSBA \cite{li2021invisible}        & 47.44 & 40.52 & 1.86  & 1.80          & 75.67 & 12.40 & 90.96          & \textbf{1.27} & \textbf{99.95} & 5.55          \\
Trojan \cite{Trojannn}             & 92.64 & 37.72 & 50.52 & 86.10         & 99.99 & 11.50 & \textbf{99.89} & 10.07         & 99.28          & \textbf{3.41} \\
WaNet \cite{nguyen2021wanet}       & 50.12 & 39.91 & 17.57 & 14.70         & 0.86  & 8.90  & 99.72          & 10.14         & \textbf{99.79} & \textbf{6.91} \\ \midrule
Average                            & 56.89 & 39.10 & 21.77 & 22.66         & 51.26 & 9.94  & 79.65          & 8.28          & \textbf{99.68} & \textbf{5.52} \\ \bottomrule
\end{tabular}}
\end{table}
\subsubsection{Further extension to test time detection}
%intro
In this work, we demonstrate that the CNPD framework can be effectively extended to detect poisoned samples during the inference phase. 
Specifically, we introduce a poisoned sample detection mechanism that compares the network's output with and without the NP layer. By evaluating any discrepancies between the two outputs, we can identify whether a query sample is poisoned. Notably, this detection method is highly efficient and incurs only a negligible increase in inference cost compared to a single forward pass.

Our approach leverages the unique properties of the NP layer. When a backdoor attack is successful, the attacked sample is misclassified as the target label. However, the NP layer suppresses the backdoor feature, altering the network's output label for the poisoned sample, while the output for clean samples remains unchanged. This distinction allows us to design a simple yet effective poisoned sample detection method, as described by the following rule:
\begin{equation}
    R(\x;f_{\w},\hat{f}_{\vtheta})= \begin{cases}\text {1}, & \text { if } f_{\w}(\x)\neq \hat{f}_{\vtheta}(\x) \\ 0 , & \text { if } f_{\w}(\x)= \hat{f}_{\vtheta}(\x)\end{cases},
\end{equation}
where $\x$ is input, $1$ denotes that $\x$ is detected as a backdoor sample.
Unlike existing backdoor detection methods, this approach eliminates the need for repeated model queries, additional computational overhead, or extra training for optimization or estimation. In essence, it performs backdoor detection with only minimal computational cost beyond the initial inference. We denote this detection method as \textit{NPDT}

% 对比方法
Here, we compare the proposed NPDT with four SOTA backdoor detection methods, \ie, 
SCALE-UP \cite{guoscale}, SentiNet \cite{chou2020sentinet}, STRIP \cite{gao2019strip}, and IBD-PSC \cite{hou2024ibdpsc}.
% metric
 For the comparison of detection performance, we evaluate the True Positive Rate (\textbf{TPR}) and False Positive Rate (\textbf{FPR}) at a fixed threshold for each detection method. Superior detection performance is indicated by a higher TPR and a lower FPR.
Tab. \ref{table5} shows the detection performance of our NPDT method in identifying samples during the inference phase. Compared to existing detection methods, our approach demonstrates a very high TPR of 99.68\% on average and a very low FPR of 5.52\%. The FPR is not zero due to the inherent limitations in the model’s classification accuracy. However, the exceptionally high TPR highlights the outstanding performance of our method against malicious Attacks.

%%%%%%

\subsubsection{Visualization on purified features}
In this work, we mitigate backdoors by filtering the intermediate poisoned features of the network through a learnable neural polarizer (NP) layer. To further understand the impact of this mechanism, we conduct a detailed analysis of feature changes before and after passing through the NP layer. We perform two visualization experiments as follows.

\begin{figure*}[h]
    \centering
    \includegraphics[width=0.8\linewidth]{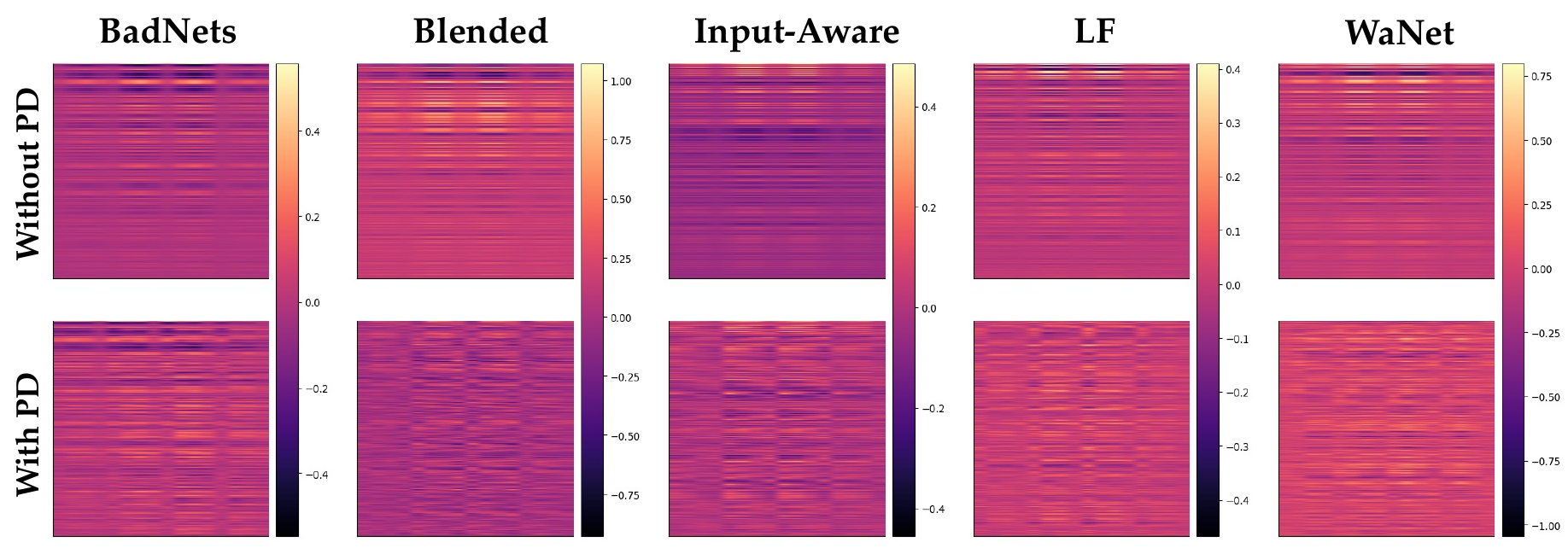}
    \caption{Feature maps of poisoned samples in the final convolutional layer of PreAct-ResNet18 network, with and without the neural polarizer layer, on CIFAR-10 dataset.}
    \label{feature_map}
\end{figure*}

\textbf{Visualization of feature map changes before and after the NP layer.}
An effective NP layer should suppress features that exhibit high activation values on backdoor-related neurons. Here, we use the TAC metric \cite{zheng2022data} to measure the correlation between neurons and the backdoor effect. We then rank these neurons according to their correlation with the backdoor and visualize the feature maps of poisoned samples on these neurons. We compare the feature maps between backdoored models with the defense models, as shown in Fig. \ref{feature_map}. Note that we use the output of the final convolutional layer of the network for visualization. In each subplot in the figure, the neurons are arranged from top to bottom according to their correlation with the backdoor, from high to low. As observed, in the backdoored model (\textbf{first row}), poisoned samples exhibit highlighted activations on the backdoor-related neurons, whereas in the defense model (\textbf{second row}), poisoned samples no longer show highlights on the backdoor-related neurons. This indicates that the backdoor-related features have successfully been suppressed.

\textbf{Visualization of feature space changes between poisoned and clean samples.}
To further compare NP's effects on clean and poisoned samples, we analyze it from the perspective of feature subspaces. Specifically, we examine the following four types of features: 1) Features of clean samples under backdoored models (\textbf{Clean before}); 2) Features of clean samples under defense models (\textbf{Clean after}); 3) Features of poisoned samples under backdoored models (\textbf{BD before}); 4) Features of poisoned samples under defense models (\textbf{BD after}).

\begin{figure*}[h]
    \centering
    \includegraphics[width=0.8\linewidth]{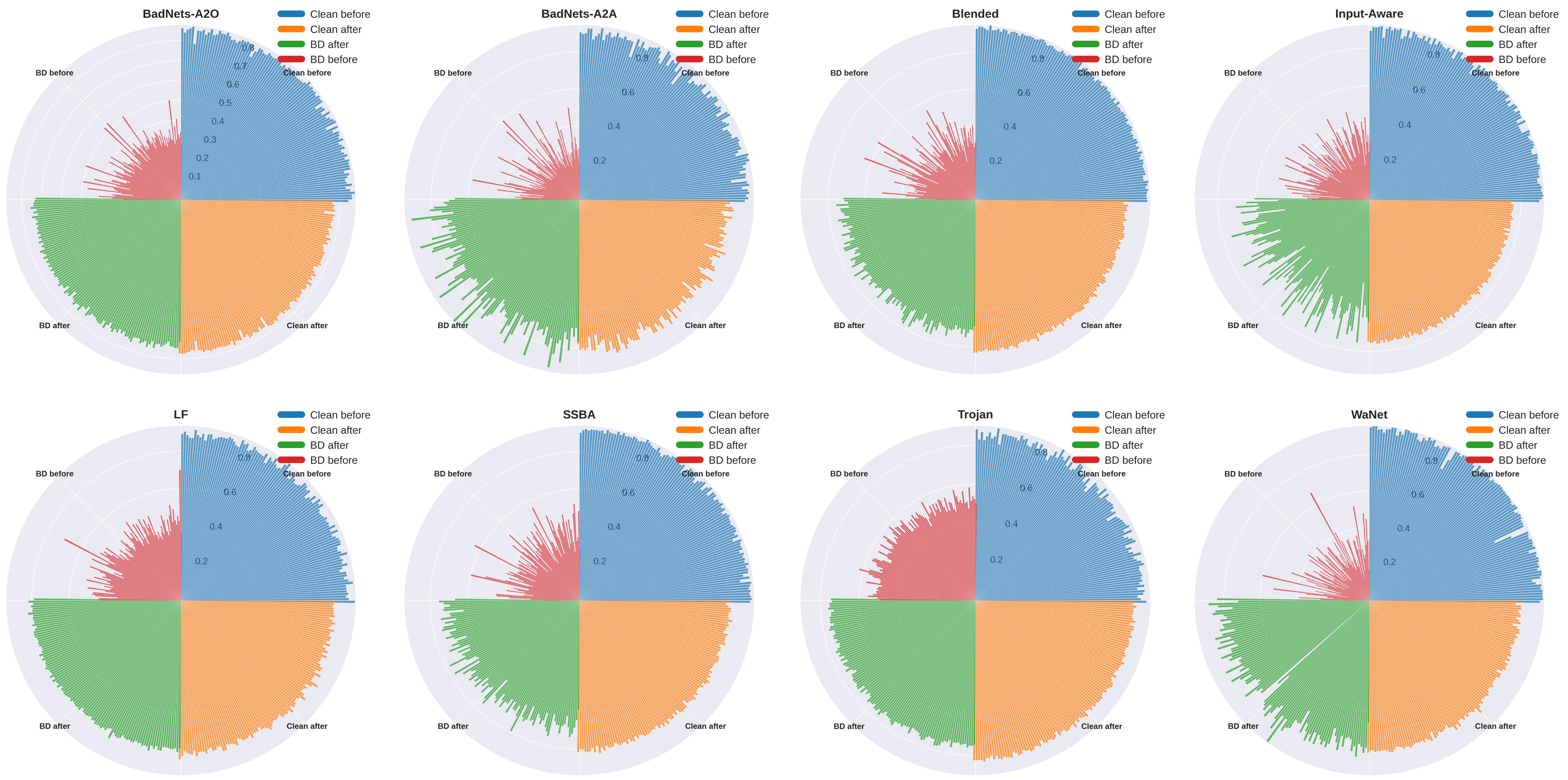}
    \caption{Visualization of clean and poisoned samples with and without the neural polarizer layer on CIFAR-10 dataset.}
    \label{visual_bar}
\end{figure*}

We perform Singular Value Decomposition (SVD) \cite{stewart1993early} on the “Clean before” and “BD before” groups, extracting two principal component vectors, \( v_{\text{clean}} \) and \( v_{\text{bd}} \). For each of the four sample groups, we compute the normalized similarity for the two vectors. The results are visualized as the components along \( v_{\text{clean}} \), as shown in Fig. \ref{visual_bar}. It can be observed that:  
\begin{itemize}
    \item After passing through the NP layer ("Clean after," orange bars), the feature activations remain largely aligned with \( v_{\text{clean}} \), suggesting that the NP layer preserves clean sample features effectively while applying minimal perturbations.
    \item The "BD after" group (green bars) exhibits strong activations on \( v_{\text{clean}} \), and suppressed activations on \( v_{\text{bd}} \), different from the "BD before" group (red bars).
\end{itemize}
The visualization demonstrates that the NP layer successfully suppresses backdoor-related feature activations while retaining the integrity of clean sample features. This selective filtering mechanism confirms the robustness of the NP layer against various types of backdoor attacks.

\begin{table*}[]
\caption{Running time of different defense methods with CIFAR-10 dataset on PreActResNet18.}
\label{table12}
\renewcommand\arraystretch{1.1}
\centering
\setlength{\tabcolsep}{4pt}
\scalebox{0.99}{%
\begin{tabular}{c|cccccccccccccc}
\toprule
Defense (Min.)$\Rightarrow$ &
  FP  &
  NAD &
  NC  &
  ANP  &
  i-BAU &
  EP  &
  FT-SAM &
  SAU &
  FST  &
  \textbf{NPD}  &
  \textbf{r-CNPD} &
  \textbf{e-CNPD} &
  \textbf{a-CNPD} \\ \midrule
  CIFAR-10 &
  12.22 &
  1.68 &
  25.38 &
  7.45 &
  2.82 &
  2.33 &
  10.85 &
  20.33 &
  1.55 &
  4.75 &
  36.6 &
  3.5 &
  3.35 \\
Tiny ImageNet &
  55.7 &
  4.82 &
  845.98 &
  28.2 &
  14.78 &
  5.03 &
  31.53 &
  22.18 &
  7.33 &
  7.17 &
  503.87 &
  7.48 &
  8.5 \\ \bottomrule
\end{tabular}}
\end{table*}

\subsubsection{Running time comparison}
Here, we demonstrate the training efficiency of our method. We tested the training times of all defense methods on two datasets, namely CIFAR-10 and Tiny ImageNet, with the results presented in Tab. \ref{table12}. As observed, the r-CNPD method incurs a significant training cost, while our NPD, e-CNPD and a-CNPD methods are significantly more efficient compared to existing defense approaches.

\section{Conclusion\label{sec5}}
In conclusion, we have proposed a novel and lightweight backdoor defense approach, Neural Polarizer Defense (NPD), which effectively filters out malicious triggers from poisoned samples while preserving benign features. Drawing inspiration from the concept of optical polarizers, NPD integrates a learnable neural polarizer (NP) as an intermediate layer within compromised models, effectively mitigating backdoor attacks while maintaining benign performance. 
Furthermore, to enhance the defense capabilities, we introduce class-conditional neural polarizer-based defense (CNPD), which emulates the targeted adversarial approach by combining class information predicted by the backdoored model with the features to be purified. 
Through three distinct instantiations of CNPD—r-CNPD, e-CNPD, and a-CNPD—we show how to balance backdoor mitigation with computational efficiency, especially for large class numbers.
Moreover, we provide a theoretical guarantee for the class-conditional neural polarizer, establishing an upper bound on backdoor risk.
Our method also includes an effective poisoned sample detection mechanism during inference, requiring minimal additional computational cost while achieving high detection accuracy. Extensive experiments across diverse neural network architectures and datasets show that our approach outperforms existing backdoor defense techniques, offering a robust and efficient solution with minimal reliance on clean training data.
Overall, the proposed neural polarizer offers a promising direction for enhancing backdoor defense in deep learning models, balancing attack mitigation with performance preservation.

\textbf{Limitations and future work.}
Although minimal, the method still relies on some clean training data for effective NP layer training. In scenarios with no clean data, defense performance may be compromised. Therefore, a promising direction for future work is to explore data-free training or training the neural polarizer with out-of-distribution samples. Applying our method to multimodal models or large-scale neural networks could be a promising direction, as full finetuning these models is often computationally expensive. By integrating our defense mechanism with such models, we could achieve more robust and scalable backdoor defenses.

\section*{Acknowledgments}
Baoyuan Wu is supported by the Guangdong Basic and Applied Basic Research Foundation (No. 2024B1515020095), 
Shenzhen Science and Technology Program (No. RCYX20210609103057050 and JCYJ20240813113608011), 
Sub-topic of Key R\&D Projects of the Ministry of Science and Technology (No. 2023YFC3304804), 
Longgang District Key Laboratory of Intelligent Digital Economy Security, and Guangdong Provincial Special Support Plan - Guangdong Provincial Science and Technology Innovation Young Talents Program (No. 2023TQ07A352). 
Hongyuan Zha is supported in part by the Shenzhen Key Lab of Crowd Intelligence Empowered Low-Carbon Energy Network (No.
ZDSYS20220606100601002).

\bibliographystyle{plain}
\bibliography{main}
% \input{sections/7_bio}

%arabic 阿拉伯数字
%roman 小写的罗马数字
%Roman 大写的罗马数字
%alph 小写字母
%Alph 大写字母

% \textbf{Organization of the Supplementary Materials.}
% We have provided the table of contents below to facilitate easy navigation of the Supplementary Materials.
% \begin{itemize}
%     \item Section \ref{A} presents the detailed algorithms for the proposed e-CNPD and a-CNPD, alongside an introduction to the Projected Gradient Descent (PGD) method. Additionally, we provide the detailed proof of our Theorem \ref{theorem1} and \ref{theorem2} in the main manuscript.
%     \item Section \ref{B} covers the implementation details, including the datasets, the attack and defense methods we compare, and our proposed approaches.
%     \item Section \ref{C} displays more defense results, including the results on Tiny ImageNet dataset, as well as on VGG19-BN architecture.
% \end{itemize}

% \setcounter{section}{0}
% \setcounter{equation}{0}
% \setcounter{page}{1}
\clearpage
% \twocolumn[
% \begin{center}
%     \begin{spacing}{2}
%         {\LARGE \textbf{Supplementary Materials of ``Conditional Neural Polarizer: Leveraging Backdoor Models for Self-Purification''}}
%         \vspace{1cm}
%     \end{spacing}
% \end{center}
% ]

\long\def\comment#1{}
\renewcommand\thesection{\Alph{section}}
% \appendix
\section*{\large Appendix}
We have provided the table of contents below to facilitate easy navigation of the Appendix.
\begin{itemize}
    \item Section \ref{app_A} presents the detailed algorithms for the proposed e-CNPD and a-CNPD, alongside an introduction to the Projected Gradient Descent (PGD) method. Additionally, we provide the detailed proof of our Theorem \ref{theorem1} and \ref{theorem2} in the main manuscript.
    \item Section \ref{app_B} covers the implementation details, including the datasets, the attack and defense methods we compare, and our proposed approaches.
    \item Section \ref{app_C} displays more defense results, including the results on Tiny ImageNet dataset, as well as on VGG19-BN architecture.
\end{itemize}

\setcounter{section}{0}

\section{More Algorithmic Details and Theoretical result\label{app_A}}
\subsection{Introduction to PGD Algorithm}
In this section, we introduce the PGD attack algorithm \cite{madry2017towards}, which plays a crucial role in three of our defense methods.

Projected Gradient Descent (PGD) is a multi-step extension of the Fast Gradient Sign Method (FGSM) \cite{kurakin2018adversarial}. It performs projected gradient descent on the negative loss function. Specifically, given an input \((\x, y)\) and a model \(\hat{f}_{\vtheta}\), the adversarial example is computed iteratively as follows:

\begin{equation} 
\x^{n+1} = \Pi_{\x + \mathcal{S}}\left(\x^n + \alpha \operatorname{sgn}\left(\nabla_{\x} L_{\text{CE}}(\hat{f}_{\vtheta}(\x), y)\right)\right),
\end{equation}
where \(\Pi_{\x + \mathcal{S}}\) denotes the projection onto the \(\rho\)-ball with \(L_p\) norm: \(\|\x^{n+1} - \x\|_p \leq \rho\), and \(\x^0 = \x\), \(n = 0, \cdots, N-1\). Here, \(\alpha\) is the step size.

However, in Section 3 of the main manuscript, we utilize a confidence-guided targeted PGD approach to generate the targeted adversarial example. In this case, the adversarial example is updated as follows:

\begin{equation} \label{eq:pgd}
\x^{n+1} = \Pi_{\x + \mathcal{S}}\left(\x^n - \alpha \operatorname{sgn}\left(\nabla_{\x} L_{\text{CE}}(\hat{f}_{\vtheta}(\x), T)\right)\right),
\end{equation}
where \(T\) denotes the target label for \(\x\). For convenience, we define the adversarial perturbation as:

\begin{equation} \label{eq:pert}
\boldsymbol{\delta} = \x^{n+1} - \x^n.
\end{equation}

\subsection{Detailed Algorithms of e-NPD and a-CNPD}
In the main manuscript, we introduce the algorithm for CNPD. Here, we provide the detailed algorithms for e-CNPD and a-CNPD, as shown in Alg. \ref{alg2_enpd} and \ref{alg3_anpd}.

\begin{algorithm}[h]
\caption{Embedding based Class-conditional NPD (e-CNPD)}\label{alg2_enpd}
\begin{algorithmic}[1]
\STATE \textbf{Input:} Training set $\mathcal{D}_{bn}$, backdoored model $f_{\w}$, neural polarizer $\hat{g}_{\vTheta}$, class embedding $e(c),c\in {1,\ldots,C}$,
learning rate $\eta>0$, perturbation bound $\rho>0$, norm $p$, hyper-parameters $\lambda_1$,$\lambda_2$,$\lambda_3>0$, training epochs $\mathcal{T}$, number of PGD steps $N$.
\STATE \textbf{Output:} Model $f(\w,\vTheta)$.
\STATE Initialize $g_{\vtheta}$. Fix $\w$, and construct the composed network $f(\w,\vTheta)$.
\FOR {$t=0,...,$ $\mathcal{T} -1$}
\FOR {mini-batch $\mathcal{B}=\{(\x_i,y_i)\}_{i=1}^{b}\subset \mathcal{D}_{bn}$}
\STATE Randomly sample target labels $\{y'_i\neq y_i|y'_i\in \{1,\ldots,C\}\}_{i=1}^{b}$;
\FOR{$n=0,...,$ $N -1$}
\STATE Generate perturbations $\{(\boldsymbol{\delta_i})\}_{i=1}^{b}$ with $\boldsymbol{\|\delta_{i}\|_p}\leq \rho$ and target labels $\{y'_i\}_{i=1}^{b}$ by targeted PGD attack \cite{madry2017towards} via Eq. (\ref{eq:at});
\ENDFOR
\STATE Select the corresponding NPs for clean batches $\{\x_i\}_{i=1}^{b}$ and perturbed batches $\{\x_i+\boldsymbol{\delta_{i}}\}_{i=1}^{b}$ according to labels $\{y_i\}_{i=1}^{b}$ and target labels $\{y'_i\}_{i=1}^{b}$ respectively. 
\STATE Update $\vTheta$ via outer minimization of Eq. (\ref{loss}) by SGD.
\ENDFOR
\ENDFOR
\RETURN Model $f(\w,\vTheta)$. 
\end{algorithmic}
\end{algorithm}

\begin{algorithm}[h]
\caption{Attention-based Class-conditional NPD (a-CNPD)}\label{alg3_anpd}
\begin{algorithmic}[1]
\STATE \textbf{Input:} Training set $\mathcal{D}_{bn}$, backdoored model $f_{\w}$, neural polarizer $\hat{g}_{\vTheta}$, class embedding $e(c),c\in {1,\ldots,C}$,
learning rate $\eta>0$, perturbation bound $\rho>0$, norm $p$, hyper-parameters $\lambda_1$,$\lambda_2$,$\lambda_3>0$, training epochs $\mathcal{T}$, number of PGD steps $N$.
\STATE \textbf{Output:} Model $f(\w,\vTheta)$.
\STATE Initialize $g_{\vtheta}$. Fix $\w$, and construct the composed network $f(\w,\vTheta)$.
\FOR {$t=0,...,$ $\mathcal{T} -1$}
\FOR {mini-batch $\mathcal{B}=\{(\x_i,y_i)\}_{i=1}^{b}\subset \mathcal{D}_{bn}$}
\STATE Randomly sample target labels $\{y'_i\neq y_i|y'_i\in \{1,\ldots,C\}\}_{i=1}^{b}$;
\FOR{$n=0,...,$ $N -1$}
\STATE Generate perturbations $\{(\boldsymbol{\delta_i})\}_{i=1}^{b}$ with $\boldsymbol{\|\delta_{i}\|_p}\leq \rho$ and target labels $\{y'_i\}_{i=1}^{b}$ by targeted PGD attack \cite{madry2017towards} via Eq. (\ref{eq:at});
\ENDFOR
\STATE Select the corresponding NPs for clean batches $\{\x_i\}_{i=1}^{b}$ and perturbed batches $\{\x_i+\boldsymbol{\delta_{i}}\}_{i=1}^{b}$ according to labels $\{y_i\}_{i=1}^{b}$ and target labels $\{y'_i\}_{i=1}^{b}$ respectively. 
\STATE Update $\vTheta$ via outer minimization of Eq. (\ref{loss}) by SGD.
\ENDFOR
\ENDFOR
\RETURN Model $f(\w,\vTheta)$.
\end{algorithmic}
\end{algorithm}
%%%

\subsection{Proof of Theorem 1}
% In this section, we provide the proof of our theorem 1 and 2.

% \begin{thm}
% Assume that $\mathbb{P}(m=0)\in (0,1)$ and $\phi_{XM}(\x_i,m)\neq \phi_{XM}(\x_j,m)$ if $\x_i\neq \x_j$. Given a poisoned model $h_{bd}$ trained by minimizing (\ref{mse}), there exists a non-trivial linear projection operator $P$ such that
% $$\text{Cov}(\langle \hat{\phi}_{XM}(\x,m), h_{bd}\rangle_{\mathcal{H}_{XM}}, m)=0,$$
% where $\hat{\phi}_{XM}(\x,m) = P\phi_{XM}(\x,m)$ is the projected feature of $\phi_{XM}(\x,m)$ and $\langle \cdot, \cdot\rangle_{XM}$ is the inner product in $\gH_{XM}$.
% \label{theorem1}
% \end{thm}

\begin{proof}
    To prove Theorem 1, we consider two cases:
    
    \noindent \textbf{Case 1: $\text{Cov}(\phi_{X}(\bm{X})|Y=y,M(\bm{X})|Y=y) = 0$:} This case contains either exclusively clean ($P(M(\bm{X})=1|Y=y)=0$) or entirely poisoned ($P(M(\bm{X})=1|Y=y)=1$) samples in class $y$, or the feature mapping $\phi_{X}$ fail to capture the correlation between $\bm{X}$ and its poisoning status for samples in class $y$. In such case, $\text{Cov}(\langle \phi^y_{X}(\bm{X}), h_{bd}\rangle_{\mathcal{H}_{X}}|Y=y, M(\bm{X})|Y=y)=0$ holds for any non-trivial projectors by the property of covariance operator with linear transformation.

    \noindent \textbf{Case 2: $\text{Cov}(\phi_{X}(\bm{X})|Y=y,M(\bm{X})|Y=y) \neq 0$:} In this case, we first recall that for a real variable $\bm{X}$, a binary variable $M(\bm{X})$, and kernel space $\mathcal{H}_{X}$, under the assumption $\phi_{X}(\x_i)\neq \phi_{X}(\x_j)$ if $\x_i\neq \x_j$ and $\text{Cov}(\phi_{X}(\bm{X}),M(\bm{X})) \neq 0$,  \cite{wei2023mean} and \cite{zhu2024neural} say that there exists a subspace $\mathcal{H}_{sub}$ of $\mathcal{H}_{X}$, in which each function $h\in\mathcal{H}_{sub}$ satisfies $\text{Cov}(h(\bm{X}),M(\bm{X}))=0$. And a projection operator $P$ from $\mathcal{H}_{X}$ to $\mathcal{H}_{sub}$ can be constructed by the eigenfunctions of the operator $\Sigma_{XM(\bm{X})}\Sigma_{M(\bm{X})\bm{X}}$ where $\Sigma_{\bm{X}M(\bm{X})}$ is the covariance operator between $\phi_{X}(\x)$ and $\phi_{M}(M(\x))$ for  polynomial kernel $\phi_M$ with degree 2.

    Then, conditioning on $Y=y$, the above analysis can be extended to a subspace $h_{sub}^y$ with $\text{Cov}(h(\bm{X})|Y=y,M(\bm{X})|Y=y)=0$ for each $h\in h_{sub}^y$ and a projection operator $P_y$ can be constructed by the eigenfunctions of 
    $\Sigma_{\bm{X}M(\bm{X})|Y=y}\Sigma_{M(\bm{X})\bm{X}|Y=y}$ where $\Sigma_{\bm{X}M(\bm{X})|Y=y}$ is the class-conditional covariance operator. In such case, a non-zero class-conditional covariance between $\phi_{X}(\x)$ and $M(\x)$ dictates a specific, non-trivial projection operator $P_y$. Such projection operator can be applied to project the feature $\phi_{X}(\x)$ and eliminate the correlation between the prediction of $\x$ and its poisoning status.
\end{proof}

\subsection{Proof of Theorem 2}
% \begin{thm}
% Assume that $h_{bd}$ is a fully backdoored model, \ie, $h_{bd}(\vx_{\Delta})=T$ for all $\vx\in\mathcal{D}$. Furthermore, assume that $\|\vx_{\Delta}-\x\|_p\leq \rho$. Then, the following inequality holds:
% \begin{equation}
%     \mathcal{R}_{bd}(h)\leq \mathcal{R}_{cnpd}(h).
% \end{equation}
% \label{theorem2}
% \end{thm}

\begin{proof}
Assume that $h_{bd}$ is a fully backdoored model, \ie, $h_{bd}(\vx_{\Delta})=T$ for all $\vx\in\mathcal{D}$. Furthermore, assume that $\|\vx_{\Delta}-\x\|_p\leq \rho$. 

Then, we have 

\begin{align}
        & \mathcal{R}_{cnpd}(h) \\&=\frac{\sum\limits_{\x\sim\mathcal{D},t\in[1,C]}\max\limits_{\|\delta\|_{p}\leq\rho}\mathbb{I}(h_{bd}(\x_{\Delta})=t,t\neq y)\mathbb{I}(h(\x+\vdelta)=t)}{|\mathcal{D}|}\\
        &=\frac{\sum\limits_{\x\sim\mathcal{D}}\max\limits_{\|\delta\|_{p}\leq\rho}\mathbb{I}(h(\x+\vdelta)=T)}{|\mathcal{D}|}\\
        &\geq \frac{\sum\limits_{\x\sim\mathcal{D}}\mathbb{I}(h(\x_{\Delta})=T)}{|\mathcal{D}|}.
\end{align}
\end{proof}

\section{More implementation details\label{app_B}}
In this section, we provide additional implementation details, including an overview of the attacks compared, specifics of their implementation, an introduction to the defenses considered, and the implementation of the methods we propose. All experiments are conducted on one RTX 3090Ti GPU.

\subsection{Introduction of Attacks} 
\begin{figure}[h]
    \centering
    \includegraphics[width=\linewidth]{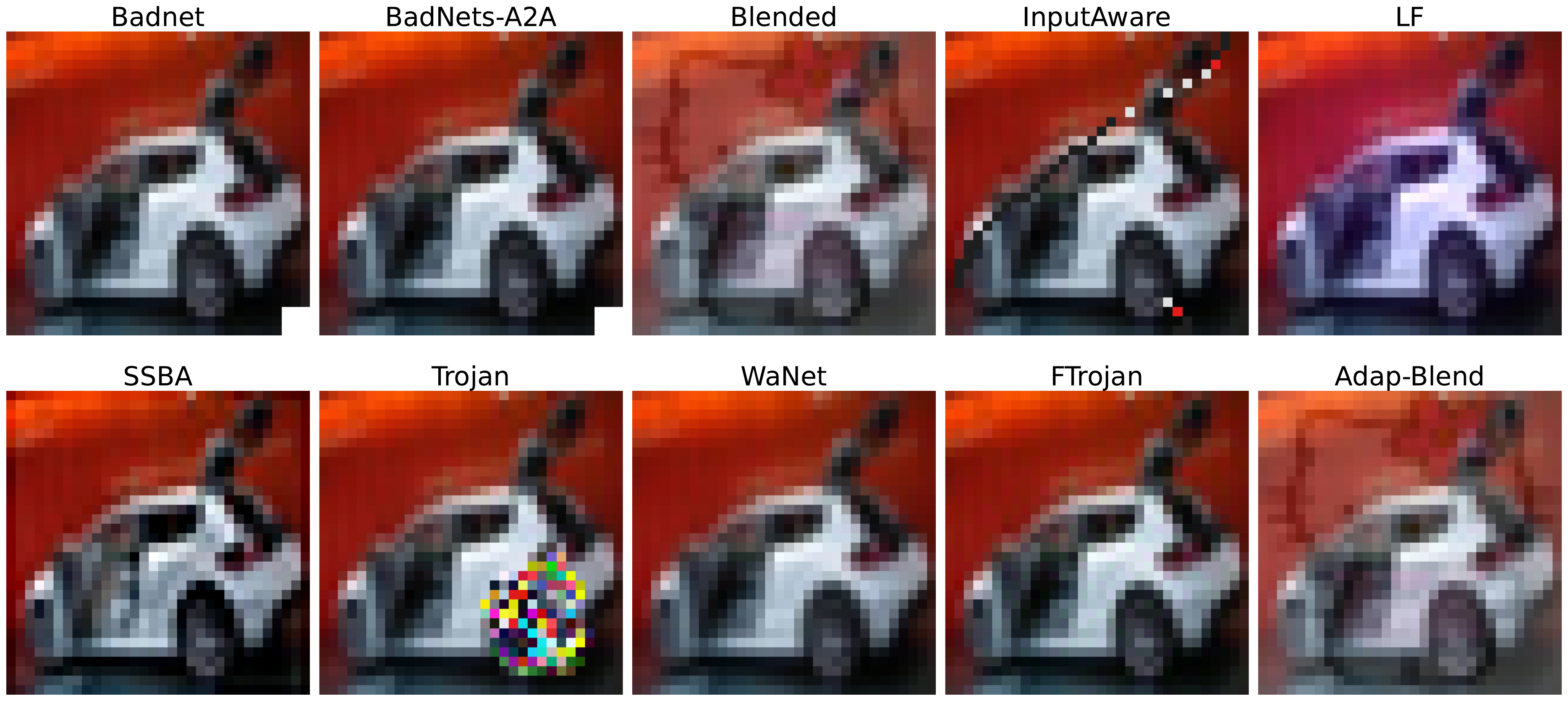}
    \caption{Visualization of poisoned samples under different backdoor attacks on CIFAR-10 dataset.}
    \label{visual_trigger_cifar}
\end{figure}

\begin{figure}[h]
    \centering
    \includegraphics[width=\linewidth]{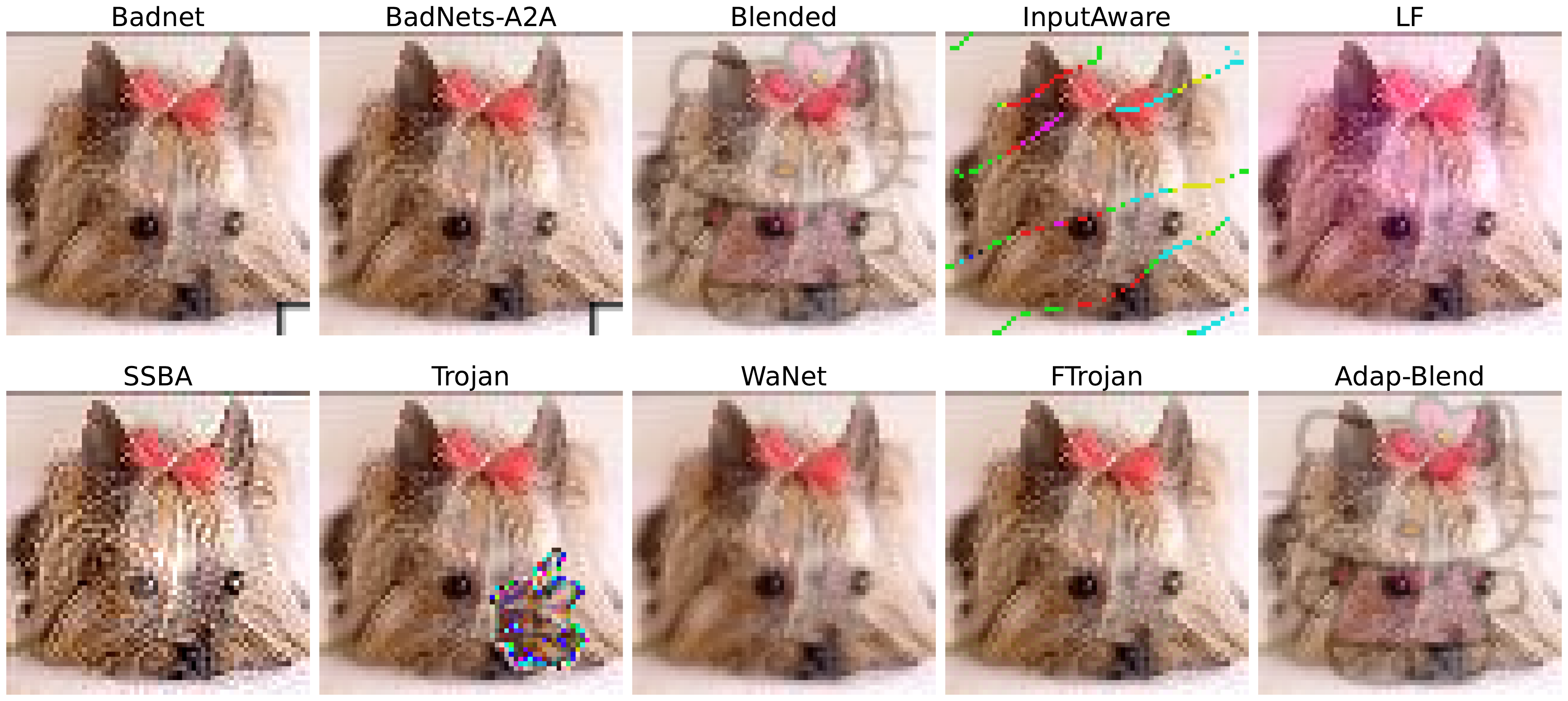}
    \caption{Visualization of poisoned samples under different backdoor attacks on Tiny ImageNet dataset.}
    \label{visual_trigger_tiny}
\end{figure}

Fig. \ref{visual_trigger_cifar} and \ref{visual_trigger_tiny}  illustrates visual representations of poisoned samples for various backdoor attacks, evaluated on the CIFAR-10 and Tiny ImageNet datasets. 
We follow BackdoorBench \cite{wubackdoorbench} to categorize the ten backdoor attacks according to various criteria as in Tab. \ref{app_table1}. We also follow the attack configuration as that in BackdoorBench.

% \usepackage{multirow}
% \usepackage{graphicx}
\begin{table*}[!h]
\centering
\caption{Categories of the ten backdoor attack methods.}
\resizebox{\textwidth}{!}{%
\begin{tabular}{c|cc|cc|cc|cc|cc}
\toprule
\multirow{2}{*}{ATTACK} &
\multicolumn{2}{c|}{Fusion pattern} &
\multicolumn{2}{c|}{Size of trigger} &
\multicolumn{2}{c|}{Visibility of trigger} &
\multicolumn{2}{c|}{Variability of trigger} &
\multicolumn{2}{c}{Num of target classes} \\ \cline{2-11} 
& Additive   & Non-Additive & Local      & Global     & Visible    & Invisible  & agnostic   & speciﬁc    & All2one    & All2all    \\ \midrule
BadNets-A2O \cite{gu2019badnets} & \checkmark &              & \checkmark &            & \checkmark &            & \checkmark &            & \checkmark &            \\
BadNets-A2A \cite{gu2019badnets} & \checkmark &              & \checkmark &            & \checkmark &            & \checkmark &            &            & \checkmark \\
Blended \cite{chen2017targeted}     &    \checkmark         &   &            & \checkmark & \checkmark &            & \checkmark &            & \checkmark &            \\
Input-Aware \cite{nguyen2020input} & \checkmark &              & \checkmark &            & \checkmark &            &            & \checkmark & \checkmark &            \\
LF \cite{zeng2021rethinking}          &            & \checkmark   &            & \checkmark &            & \checkmark &            & \checkmark & \checkmark &            \\
SSBA \cite{li2021invisible}        &            & \checkmark   &            & \checkmark &            & \checkmark &            & \checkmark & \checkmark &            \\
Trojan \cite{Trojannn}      & \checkmark &              & \checkmark &            & \checkmark &            & \checkmark &            & \checkmark &            \\
WaNet \cite{nguyen2021wanet}       &            & \checkmark   &            & \checkmark &            & \checkmark &            & \checkmark & \checkmark &            \\ 
FTrojan \cite{wang2022invisible}      &  &     \checkmark         &  &  \checkmark          &  &    \checkmark        & \checkmark &            & \checkmark &            \\
Adap-Blend \cite{qi2023revisiting}      &    \checkmark         &   &            & \checkmark & \checkmark &            & \checkmark &            & \checkmark &            \\
\bottomrule
\end{tabular}%
\label{app_table1}
}
\end{table*}

Specifically, the details of these attacks are as follows:
\begin{itemize}
    \item \textbf{BadNets} \cite{gu2019badnets}: BadNets is a trigger-additive attack that inserts a fixed patch (a $3 \times 3$ white square for CIFAR-10 and GTSRB, and $6 \times 6$ for Tiny ImageNet in our paper, which is consistent with BackdoorBench.) into the image, and the corresponding label is altering to the target label.
    
    \item \textbf{Blended Backdoor Attack} (Blended) \cite{chen2017targeted}: The Blended attack involves merging a predefined image (Hello Kitty, in our work) with the original image using a blend coefficient $\alpha = 0.2$, as per the BackdoorBench standard.

    \item \textbf{Input-Aware Dynamic Backdoor Attack} (Input-Aware) \cite{nguyen2020input}: This attack is controllable during training. It first trains a trigger generator using adversarial methods. The generator then creates specific triggers for each sample during the model's training phase.

    \item \textbf{Low Frequency Attack} (LF) \cite{zeng2021rethinking}: The LF attack begins by learning a universal adversarial perturbation (UAP) while filtering out high-frequency components. The remaining low-frequency perturbation is then applied to clean samples to generate poisoned data.

    \item \textbf{Sample-Specific Backdoor Attack} (SSBA) \cite{li2021invisible}: SSBA utilizes an autoencoder, which is first trained. The autoencoder is then used to combine a trigger with the clean samples, producing poisoned images.

    \item \textbf{Trojan Backdoor Attack} (Trojan) \cite{Trojannn}: Similar to the LF attack, Trojan also generates a universal adversarial perturbation (UAP). This perturbation is then applied to clean samples to create poisoned ones.

    \item \textbf{Warping-Based Poisoned Networks} (WaNet) \cite{nguyen2021wanet}: WaNet employs a warping function that perturbs the clean samples to generate poisoned versions. During training, the adversary controls the process to ensure the model learns this specific warping transformation.
    \item \textbf{Trojaning attack the frequency domain} (Ftrojan) \cite{wang2022invisible}: Ftrojan introduces adversarial perturbations in the frequency domain, such that the poisoned image are visually imperceptible.
    \item \textbf{Adaptive Backdoor Poisoning Attack} (Adap-Blend) \cite{qi2023revisiting}: Adap-Blend designs an asymmetric trigger planting strategy that weakens triggers for data poisoning to diversify the latent representations of poisoned samples, and uses a stronger trigger during test time to improve attack success rate. 
\end{itemize}

\subsection{Introduction of Defenses} 
In this work, we compare our methods with ten SOTA backdoor defense methods. We conduct defense experiments following BackdoorBench. The detailed introduction of these methods are as follows:
\begin{itemize}
    \item \textbf{Fine-Pruning (FP)} \cite{liu2018fine}: FP first prunes neurons that remain dormant when processing benign samples. Then, it fine-tunes the pruned model to recover its utility and improve overall performance.
    
    \item \textbf{Neural Attention Distillation (NAD)} \cite{li2021neural}: NAD employs a knowledge distillation strategy that aims to distill clean knowledge from a potentially contaminated model using clean samples.
    \item \textbf{Neural Cleanse (NC)} \cite{wang2019neural}: NC detects backdoors by searching for a minimal Universal Adversarial Perturbation (UAP). Once a backdoor is detected, it purifies the model by unlearning the identified UAP.
    \item \textbf{Adversarial Neuron Pruning (ANP)} \cite{wu2021adversarial}: ANP is a pruning-based method. It observes that neurons associated with backdoors are more susceptible to adversarial perturbations and employs a minimax optimization strategy to identify and mask these compromised neurons.
    \item \textbf{Implicit Backdoor Adversarial Unlearning (i-BAU)} \cite{zeng2022adversarial}: i-BAU utilizes an implicit hyper-gradient method to optimize adversarial training, removing backdoor influences by unlearning adversarial perturbations.
    \item \textbf{Entropy-based pruning (EP)} \cite{zheng2022pre}: EP identifies significant differences in the moments of pre-activation distributions between benign and poisoned data in backdoor neurons in contrast to clean neurons, which is utilized for model pruning.
    \item \textbf{Fine-Tuning with Sharpness-Aware Minimization (FT-SAM) \cite{zhu2023enhancing}}: FT-SAM employs sharpness-aware minimization to fine-tune the poisoned model, improving its robustness by minimizing sharp loss regions.
    \item \textbf{Feature Shift Tuning (FST)} \cite{min2024towards}: FST encourages feature shifts by reinitializing the linear classifier and fine-tuning the model to counteract backdoor effects.
    \item \textbf{Shared Adversarial Unlearning (SAU)} \cite{wei2023shared}: SAU generates shared adversarial examples and then unlearns these examples to purify the model, targeting adversarial behavior.
\end{itemize}

\subsection{Implementation Details} 
In this work, starting from the concept of a neural polarizer, we propose NPD and CNPD with three instantiations: r-CNPD, e-CNPD, and a-CNPD. In practical experiments, the selection of their hyperparameters differs slightly based on their performance.

First, for the generation of adversarial examples, we use a unified $l_2$ norm for adversarial perturbation generation, with a perturbation bound of 3 for the CIFAR-10 and GTSRB datasets, and 6 for the Tiny ImageNet dataset. For each training batch, the inner loop uses PGD algorithm to generate adversarial examples, with 5 iterations and learning rate 0.1 for the inner loop.

Regarding our NPD, we use the same hyperparameters as in our Conference \cite{zhu2024neural}. Specifically, the training of neural polarizer consists of 50 epochs for all three datasets, with a warm-up period of 5 epochs. We use a learning rate of 0.01 with a weight decay of 0.0005 and a momentum of 0.9. The NP layer is inserted before the third convolution layer of the fourth layer for PreAct-ResNet18.

Next, we consider CNPD. Regarding the loss function, for the e-CNPD and a-CNPD methods, the hyperparameters $\lambda_1$, $\lambda_2$, and $\lambda_3$ are set to 1, 0.4, and 0.4 for the CIFAR-10 dataset, 1, 0.5, and 0.5 for the GTSRB dataset, and 1.0, 0.1, and 0.1 for the Tiny ImageNet dataset. For the r-CNPD method, the hyperparameters $\lambda_1$, $\lambda_2$, and $\lambda_3$ are set to 1, 0.1, and 0.1 for CIFAR-10, and 1, 0.2, and 0.1 for GTSRB.

Regarding the number of training epochs, for the a-CNPD method, we train for 10, 50, and 200 epochs on the CIFAR-10, GTSRB, and Tiny ImageNet datasets, respectively. For e-CNPD, we train for 100, 50, and 100 epochs. For r-CNPD, we consistently train for 200 epochs.

In terms of the insertion layer, we find that inserting PD into either the penultimate or last convolutional layer yields good defense results. In this paper, we uniformly choose to insert it into the last convolutional layer in the main experiments for CNPD.

For optimization, we use SGD with a learning rate of 0.01, weight decay of $5 \times 10^{-4}$, and momentum of 0.9. The learning rate is set to 0.001 for e-CNPD on Tiny ImageNet dataset. 

As for the architecture, the PD structure in the r-CNPD method is identical to that of NPD, consisting of a $1 \times 1$ convolutional layer followed by a batch normalization (BN) layer. The total number of such structures is equal to the number of classes.

For the e-CNPD method, let the output features from the previous layer be denoted as $m^l(\x) \in \mathbb{R}^{c \times h \times w}$, where $c$, $h$, and $w$ represent the number of channels, height, and width, respectively. The label embedding feature for each class is of dimension $h \times w$. Thus, the feature $m^l(\x)$ corresponding to class $y$ is concatenated with the embedding $e(y)$, resulting in a feature of dimension $(c+1) \times h \times w$, which is then passed through a Conv-BN layer, followed by a ReLU activation, and subsequently processed through another Conv-BN layer to produce an output of size $c \times h \times w$.

For the a-CNPD method, the parameters include a class-specific embedding vector $e(c) \in \mathbb{R}^{c \times h}$ for each class, and three learnable matrices: linear layers $\vtheta_Q \in \mathbb{R}^{h \times h}$, $\vtheta_K \in \mathbb{R}^{h \times h}$, and a $1 \times 1$ convolutional layer $\vtheta_V$. After performing the operations as described in Eq. (\ref{eq:anpd}), the resulting feature is passed through a final Conv-BN layer to obtain the filtered output.

% \usepackage{booktabs}
\begin{table*}[t]
\centering
\caption{Defense Results on Tiny ImageNet with PreAct-ResNet18 and poisoning ratio $10.0\%$.}
\label{table3}
\setlength{\tabcolsep}{3pt} % Default value: 6pt
\scalebox{0.8}{
\begin{tabular}{c|ccc|ccc|ccc|ccc|ccc|ccc}
\toprule
Defense $\rightarrow$ & \multicolumn{3}{c|}{No Defense} & \multicolumn{3}{c|}{FP \cite{liu2018fine}} & \multicolumn{3}{c|}{NAD \cite{li2021neural}} & \multicolumn{3}{c|}{NC \cite{wang2019neural}} & \multicolumn{3}{c|}{ANP \cite{wu2021adversarial}} & \multicolumn{3}{c}{i-BAU \cite{zeng2022adversarial}} \\ 
Attack $\downarrow$ & \multicolumn{1}{c}{ACC} & \multicolumn{1}{c}{ASR} & \multicolumn{1}{c|}{DER} & \multicolumn{1}{c}{ACC} & \multicolumn{1}{c}{ASR} & \multicolumn{1}{c|}{DER} & \multicolumn{1}{c}{ACC} & \multicolumn{1}{c}{ASR} & \multicolumn{1}{c|}{DER} & \multicolumn{1}{c}{ACC} & \multicolumn{1}{c}{ASR} & \multicolumn{1}{c|}{DER} & \multicolumn{1}{c}{ACC} & \multicolumn{1}{c}{ASR} & \multicolumn{1}{c|}{DER} & \multicolumn{1}{c}{ACC} & \multicolumn{1}{c}{ASR} & \multicolumn{1}{c}{DER} \\ \midrule
BadNets-A2O \cite{gu2019badnets} & $56.12$& $99.9$&  N/A & $48.81$& $0.66$& $95.96$& $48.35$& $0.27$& $95.93$& $\cellcolor[HTML]{F1B9B6}56.12$& $99.9$& $50.00$& $50.65$& $0.12$& $97.15$& $51.63$& $95.92$& $49.74$\\
BadNets-A2A \cite{gu2019badnets} & $55.99$& $27.81$&  N/A & $47.88$& $3.19$& $58.26$& $48.29$& $2.3$& $58.91$& $\cellcolor[HTML]{F1B9B6}54.12$& $18.72$& $53.61$& $52.63$& $2.85$& $60.80$& $\cellcolor[HTML]{FEEBB8}53.52$& $12.89$& $56.22$\\
Blended \cite{chen2017targeted} & $56.49$& $99.67$&  N/A & $50.58$& $57.89$& $67.93$& $48.28$& $94.78$& $48.34$& $\cellcolor[HTML]{F1B9B6}54.50$& $96.07$& $50.80$& $43.21$& $43.80$& $71.29$& $50.76$& $95.58$& $49.18$\\
Input-Aware \cite{nguyen2020input} & $57.67$& $99.19$&  N/A & $49.18$& $3.75$& $93.48$& $50.08$& $0.61$& $95.49$& $53.28$& $0.3$& $97.25$& $\cellcolor[HTML]{F1B9B6}57.34$& $0.50$& $\cellcolor[HTML]{F1B9B6}99.18$& $53.96$& $1.29$& $97.10$\\
LF \cite{zeng2021rethinking} & $55.21$& $98.51$&  N/A & $48.18$& $63.83$& $63.83$& $49.61$& $58.01$& $67.45$& $\cellcolor[HTML]{FEEBB8}53.08$& $90.48$& $52.95$& $51.66$& $89.31$& $52.83$& $\cellcolor[HTML]{F1B9B6}53.65$& $94.27$& $51.34$\\
SSBA \cite{li2021invisible} & $55.97$& $97.69$&  N/A & $48.06$& $52.25$& $68.76$& $47.67$& $69.47$& $59.96$& $\cellcolor[HTML]{D7E8F2}52.57$& $0.05$& $\cellcolor[HTML]{FEEBB8}97.12$& $\cellcolor[HTML]{F1B9B6}54.43$& $89.63$& $53.26$& $52.39$& $84.64$& $54.73$\\
Trojan \cite{Trojannn} & $56.48$& $99.97$&  N/A & $45.96$& $8.88$& $90.28$& $48.83$& $1.01$& $95.66$& $\cellcolor[HTML]{FEEBB8}53.88$& $0.27$& $\cellcolor[HTML]{F1B9B6}98.55$& $\cellcolor[HTML]{F1B9B6}56.43$& $4.13$& $97.89$& $51.85$& $99.15$& $48.10$\\
WaNet \cite{nguyen2021wanet} & $57.81$& $96.5$&  N/A & $50.35$& $1.37$& $93.83$& $50.02$& $0.87$& $93.92$& $\cellcolor[HTML]{F1B9B6}57.81$& $96.5$& $50.00$& $\cellcolor[HTML]{FEEBB8}57.31$& $0.42$& $\cellcolor[HTML]{F1B9B6}97.79$& $53.04$& $69.82$& $60.95$\\
FTrojan \cite{wang2022invisible} & $56.0$& $100.0$&  N/A & $49.93$& $0.9$& $96.51$& $47.43$& $0.34$& $95.54$& $50.22$& $100.0$& $47.11$& $43.68$& $\cellcolor[HTML]{FEEBB8}0.00$& $93.84$& $\cellcolor[HTML]{FEEBB8}54.04$& $0.72$& $\cellcolor[HTML]{FEEBB8}98.66$\\
Adap-Blend \cite{qi2023revisiting} & $54.32$& $83.04$&  N/A & $49.43$& $3.11$& $87.52$& $48.43$& $12.86$& $82.14$& $\cellcolor[HTML]{F1B9B6}53.18$& $81.95$& $49.97$& $43.69$& $48.84$& $61.78$& $47.33$& $52.24$& $61.90$\\ \midrule
Average & $56.21$& $90.23$&  N/A & $48.84$& $19.58$& $81.64$& $48.7$& $24.05$& $79.33$& $\cellcolor[HTML]{F1B9B6}53.88$& $58.42$& $64.74$& $51.10$& $27.96$& $78.58$& $52.22$& $60.65$& $62.79$\\

\toprule

\toprule
Defense $\rightarrow$ & \multicolumn{3}{c|}{FT-SAM \cite{zhu2023enhancing}} & \multicolumn{3}{c|}{FST \cite{min2024towards}} & \multicolumn{3}{c|}{SAU \cite{wei2023shared}} & \multicolumn{3}{c|}{NPD (\textbf{Ours})} & \multicolumn{3}{c|}{e-CNPD (\textbf{Ours})} & \multicolumn{3}{c}{a-CNPD (\textbf{Ours})} \\ 
Attack $\downarrow$ & \multicolumn{1}{c}{ACC} & \multicolumn{1}{c}{ASR} & \multicolumn{1}{c|}{DER} & \multicolumn{1}{c}{ACC} & \multicolumn{1}{c}{ASR} & \multicolumn{1}{c|}{DER} & \multicolumn{1}{c}{ACC} & \multicolumn{1}{c}{ASR} & \multicolumn{1}{c|}{DER} & \multicolumn{1}{c}{ACC} & \multicolumn{1}{c}{ASR} & \multicolumn{1}{c|}{DER} & \multicolumn{1}{c}{ACC} & \multicolumn{1}{c}{ASR} & \multicolumn{1}{c|}{DER} & \multicolumn{1}{c}{ACC} & \multicolumn{1}{c}{ASR} & \multicolumn{1}{c}{DER} \\ \midrule
BadNets-A2O \cite{gu2019badnets} & $51.91$& $0.21$& $97.74$& $39.73$& $\cellcolor[HTML]{D7E8F2}0.02$& $91.74$& $\cellcolor[HTML]{D7E8F2}52.79$& $0.28$& $\cellcolor[HTML]{D7E8F2}98.14$& $49.79$& $2.51$& $95.53$& $\cellcolor[HTML]{FEEBB8}52.95$& $\cellcolor[HTML]{FEEBB8}0.00$& $\cellcolor[HTML]{F1B9B6}98.36$& $52.67$& $\cellcolor[HTML]{F1B9B6}0.00$& $\cellcolor[HTML]{FEEBB8}98.22$\\
BadNets-A2A \cite{gu2019badnets} & $52.24$& $\cellcolor[HTML]{D7E8F2}2.09$& $\cellcolor[HTML]{D7E8F2}60.98$& $45.84$& $2.92$& $57.37$& $\cellcolor[HTML]{D7E8F2}53.17$& $6.07$& $59.46$& $49.94$& $5.57$& $58.10$& $52.73$& $\cellcolor[HTML]{F1B9B6}1.08$& $\cellcolor[HTML]{F1B9B6}61.74$& $52.52$& $\cellcolor[HTML]{FEEBB8}1.61$& $\cellcolor[HTML]{FEEBB8}61.36$\\
Blended \cite{chen2017targeted} & $52.36$& $81.74$& $56.90$& $39.84$& $1.32$& $90.85$& $52.98$& $\cellcolor[HTML]{D7E8F2}0.01$& $\cellcolor[HTML]{D7E8F2}98.07$& $49.62$& $0.12$& $96.34$& $\cellcolor[HTML]{D7E8F2}53.30$& $\cellcolor[HTML]{FEEBB8}0.00$& $\cellcolor[HTML]{FEEBB8}98.24$& $\cellcolor[HTML]{FEEBB8}54.09$& $\cellcolor[HTML]{F1B9B6}0.00$& $\cellcolor[HTML]{F1B9B6}98.63$\\
Input-Aware \cite{nguyen2020input} & $52.69$& $1.01$& $96.60$& $41.15$& $\cellcolor[HTML]{F1B9B6}0.02$& $91.32$& $\cellcolor[HTML]{FEEBB8}54.68$& $0.57$& $\cellcolor[HTML]{D7E8F2}97.82$& $53.75$& $5.93$& $94.67$& $\cellcolor[HTML]{D7E8F2}54.65$& $\cellcolor[HTML]{D7E8F2}0.14$& $\cellcolor[HTML]{FEEBB8}98.02$& $54.19$& $\cellcolor[HTML]{FEEBB8}0.13$& $97.79$\\
LF \cite{zeng2021rethinking} & $51.6$& $45.15$& $74.88$& $38.62$& $2.66$& $89.63$& $50.39$& $\cellcolor[HTML]{D7E8F2}2.47$& $\cellcolor[HTML]{D7E8F2}95.61$& $49.94$& $2.48$& $95.38$& $\cellcolor[HTML]{D7E8F2}52.17$& $\cellcolor[HTML]{FEEBB8}0.01$& $\cellcolor[HTML]{F1B9B6}97.73$& $51.95$& $\cellcolor[HTML]{F1B9B6}0.00$& $\cellcolor[HTML]{FEEBB8}97.63$\\
SSBA \cite{li2021invisible} & $51.87$& $0.38$& $96.60$& $39.24$& $0.73$& $90.11$& $52.39$& $0.03$& $\cellcolor[HTML]{D7E8F2}97.04$& $49.25$& $\cellcolor[HTML]{D7E8F2}0.01$& $95.48$& $\cellcolor[HTML]{FEEBB8}52.58$& $\cellcolor[HTML]{FEEBB8}0.00$& $\cellcolor[HTML]{F1B9B6}97.15$& $51.94$& $\cellcolor[HTML]{F1B9B6}0.00$& $96.83$\\
Trojan \cite{Trojannn} & $52.28$& $0.21$& $97.78$& $41.1$& $0.58$& $92.0$& $49.99$& $\cellcolor[HTML]{D7E8F2}0.01$& $96.73$& $49.43$& $0.51$& $96.20$& $\cellcolor[HTML]{D7E8F2}53.17$& $\cellcolor[HTML]{FEEBB8}0.00$& $\cellcolor[HTML]{FEEBB8}98.33$& $52.69$& $\cellcolor[HTML]{F1B9B6}0.00$& $\cellcolor[HTML]{D7E8F2}98.09$\\
WaNet \cite{nguyen2021wanet} & $54.32$& $0.79$& $\cellcolor[HTML]{FEEBB8}96.11$& $37.71$& $\cellcolor[HTML]{FEEBB8}0.15$& $88.12$& $\cellcolor[HTML]{D7E8F2}56.13$& $41.82$& $76.50$& $52.64$& $\cellcolor[HTML]{D7E8F2}0.24$& $95.54$& $52.37$& $1.70$& $94.68$& $52.75$& $\cellcolor[HTML]{F1B9B6}0.02$& $\cellcolor[HTML]{D7E8F2}95.71$\\
FTrojan \cite{wang2022invisible} & $50.85$& $0.30$& $97.27$& $46.14$& $10.12$& $90.01$& $\cellcolor[HTML]{F1B9B6}54.38$& $\cellcolor[HTML]{D7E8F2}0.18$& $\cellcolor[HTML]{F1B9B6}99.10$& $49.55$& $2.73$& $95.41$& $52.52$& $\cellcolor[HTML]{F1B9B6}0.00$& $98.26$& $\cellcolor[HTML]{D7E8F2}52.86$& $0.23$& $\cellcolor[HTML]{D7E8F2}98.31$\\
Adap-Blend \cite{qi2023revisiting} & $50.3$& $89.73$& $47.99$& $23.6$& $\cellcolor[HTML]{D7E8F2}0.00$& $76.16$& $\cellcolor[HTML]{D7E8F2}51.88$& $2.34$& $\cellcolor[HTML]{D7E8F2}89.13$& $49.21$& $5.21$& $86.36$& $51.41$& $\cellcolor[HTML]{FEEBB8}0.00$& $\cellcolor[HTML]{FEEBB8}90.06$& $\cellcolor[HTML]{FEEBB8}52.48$& $\cellcolor[HTML]{F1B9B6}0.00$& $\cellcolor[HTML]{F1B9B6}90.60$\\ \midrule
Average & $52.04$& $22.16$& $82.29$& $39.3$& $\cellcolor[HTML]{D7E8F2}1.85$& $85.73$& $\cellcolor[HTML]{FEEBB8}52.88$& $5.38$& $90.76$& $50.31$& $2.53$& $\cellcolor[HTML]{D7E8F2}90.90$& $52.78$& $\cellcolor[HTML]{FEEBB8}0.29$& $\cellcolor[HTML]{FEEBB8}93.26$& $\cellcolor[HTML]{D7E8F2}52.81$& $\cellcolor[HTML]{F1B9B6}0.20$& $\cellcolor[HTML]{F1B9B6}93.32$\\

\bottomrule
\end{tabular}}
\end{table*}

% \usepackage{booktabs}
\begin{table*}[h]
\centering
\caption{Defense Results on CIFAR-10 with VGG19-BN and poisoning ratio $10.0\%$.}
\label{table4}
\setlength{\tabcolsep}{3pt} % Default value: 6pt
\scalebox{0.75}{
\begin{tabular}{c|ccc|ccc|ccc|ccc|ccc|ccc|ccc}
\toprule

Defense $\rightarrow$ & \multicolumn{3}{c|}{No Defense} & \multicolumn{3}{c|}{FP \cite{liu2018fine}} & \multicolumn{3}{c|}{NAD \cite{li2021neural}} & \multicolumn{3}{c|}{ANP \cite{wu2021adversarial}} & \multicolumn{3}{c|}{i-BAU \cite{zeng2022adversarial}} & \multicolumn{3}{c}{EP \cite{zheng2022pre}} \\ 
Attack $\downarrow$ & \multicolumn{1}{c}{ACC} & \multicolumn{1}{c}{ASR} & \multicolumn{1}{c|}{DER} & \multicolumn{1}{c}{ACC} & \multicolumn{1}{c}{ASR} & \multicolumn{1}{c|}{DER} & \multicolumn{1}{c}{ACC} & \multicolumn{1}{c}{ASR} & \multicolumn{1}{c|}{DER} & \multicolumn{1}{c}{ACC} & \multicolumn{1}{c}{ASR} & \multicolumn{1}{c|}{DER} & \multicolumn{1}{c}{ACC} & \multicolumn{1}{c}{ASR} & \multicolumn{1}{c|}{DER} & \multicolumn{1}{c}{ACC} & \multicolumn{1}{c}{ASR} & \multicolumn{1}{c}{DER} \\ \midrule
BadNets-A2O \cite{gu2019badnets} & $90.42$& $94.43$&  N/A & $89.11$& $12.39$& $90.37$& $86.80$& $5.77$& $92.52$& $\cellcolor[HTML]{F1B9B6}90.74$& $91.68$& $51.38$& $87.69$& $3.13$& $94.28$& $87.56$& $7.28$& $92.15$\\
BadNets-A2A \cite{gu2019badnets} & $91.16$& $84.39$&  N/A & $89.70$& $1.91$& $\cellcolor[HTML]{D7E8F2}90.51$& $88.15$& $1.6$& $89.89$& $86.02$& $1.64$& $88.80$& $86.86$& $2.19$& $88.95$& $\cellcolor[HTML]{FEEBB8}89.98$& $68.91$& $57.15$\\
Blended \cite{chen2017targeted} & $91.91$& $99.5$&  N/A & $89.60$& $93.14$& $52.02$& $87.45$& $86.98$& $54.03$& $\cellcolor[HTML]{F1B9B6}91.69$& $97.37$& $50.96$& $88.45$& $51.67$& $72.19$& $\cellcolor[HTML]{FEEBB8}90.99$& $22.62$& $87.98$\\
Input-Aware \cite{nguyen2020input} & $88.66$& $94.58$&  N/A & $\cellcolor[HTML]{F1B9B6}91.55$& $14.57$& $90.01$& $\cellcolor[HTML]{FEEBB8}91.09$& $14.06$& $90.26$& $89.67$& $36.76$& $78.91$& $89.81$& $78.93$& $57.82$& $88.92$& $3.84$& $95.37$\\
LF \cite{zeng2021rethinking} & $83.28$& $13.83$&  N/A & $88.18$& $\cellcolor[HTML]{FEEBB8}1.29$& $\cellcolor[HTML]{FEEBB8}56.27$& $85.08$& $3.07$& $55.38$& $87.74$& $5.62$& $54.11$& $87.68$& $1.47$& $56.18$& $79.36$& $18.88$& $48.04$\\
SSBA \cite{li2021invisible} & $90.85$& $95.11$&  N/A & $89.26$& $65.33$& $64.09$& $88.11$& $52.22$& $70.07$& $\cellcolor[HTML]{FEEBB8}90.15$& $\cellcolor[HTML]{D7E8F2}1.08$& $\cellcolor[HTML]{FEEBB8}96.67$& $85.61$& $12.37$& $88.75$& $\cellcolor[HTML]{D7E8F2}89.62$& $85.4$& $54.24$\\
Trojan \cite{Trojannn} & $91.57$& $100.0$&  N/A & $\cellcolor[HTML]{D7E8F2}90.04$& $29.71$& $84.38$& $87.01$& $5.17$& $95.14$& $89.68$& $\cellcolor[HTML]{D7E8F2}0.00$& $\cellcolor[HTML]{D7E8F2}99.06$& $86.4$& $2.69$& $96.07$& $89.42$& $5.23$& $96.31$\\
WaNet \cite{nguyen2021wanet} & $84.58$& $96.49$&  N/A & $\cellcolor[HTML]{F1B9B6}91.10$& $3.36$& $96.57$& $\cellcolor[HTML]{D7E8F2}90.68$& $10.23$& $93.13$& $76.23$& $\cellcolor[HTML]{F1B9B6}0.12$& $94.01$& $89.61$& $2.4$& $97.05$& $86.10$& $73.61$& $61.44$\\ \midrule
Average & $89.05$& $84.79$&  N/A & $\cellcolor[HTML]{D7E8F2}89.82$& $27.71$& $78.03$& $88.05$& $22.39$& $80.05$& $87.74$& $29.28$& $76.74$& $87.76$& $19.36$& $81.41$& $87.74$& $35.72$& $74.08$\\
\toprule
\toprule
Defense $\rightarrow$ & \multicolumn{3}{c|}{FT-SAM \cite{zhu2023enhancing}} & \multicolumn{3}{c|}{SAU \cite{wei2023shared}} & \multicolumn{3}{c|}{NPD (\textbf{Ours})} & \multicolumn{3}{c|}{r-CNPD (\textbf{Ours})} & \multicolumn{3}{c|}{e-CNPD (\textbf{Ours})} & \multicolumn{3}{c}{a-CNPD (\textbf{Ours})} \\ 
Attack $\downarrow$ & \multicolumn{1}{c}{ACC} & \multicolumn{1}{c}{ASR} & \multicolumn{1}{c|}{DER} & \multicolumn{1}{c}{ACC} & \multicolumn{1}{c}{ASR} & \multicolumn{1}{c|}{DER} & \multicolumn{1}{c}{ACC} & \multicolumn{1}{c}{ASR} & \multicolumn{1}{c|}{DER} & \multicolumn{1}{c}{ACC} & \multicolumn{1}{c}{ASR} & \multicolumn{1}{c|}{DER} & \multicolumn{1}{c}{ACC} & \multicolumn{1}{c}{ASR} & \multicolumn{1}{c|}{DER} & \multicolumn{1}{c}{ACC} & \multicolumn{1}{c}{ASR} & \multicolumn{1}{c}{DER} \\ \midrule
BadNets-A2O \cite{gu2019badnets} & $87.63$& $1.29$& $95.18$& $83.29$& $4.28$& $91.51$& $\cellcolor[HTML]{D7E8F2}89.52$& $\cellcolor[HTML]{F1B9B6}0.10$& $\cellcolor[HTML]{F1B9B6}96.72$& $88.59$& $\cellcolor[HTML]{D7E8F2}1.09$& $95.76$& $\cellcolor[HTML]{FEEBB8}89.85$& $1.96$& $\cellcolor[HTML]{D7E8F2}95.95$& $89.49$& $\cellcolor[HTML]{FEEBB8}0.64$& $\cellcolor[HTML]{FEEBB8}96.43$\\
BadNets-A2A \cite{gu2019badnets} & $88.76$& $\cellcolor[HTML]{FEEBB8}1.29$& $90.35$& $89.4$& $\cellcolor[HTML]{D7E8F2}1.52$& $\cellcolor[HTML]{FEEBB8}90.56$& $\cellcolor[HTML]{D7E8F2}89.81$& $2.36$& $90.34$& $88.06$& $3.16$& $89.07$& $\cellcolor[HTML]{F1B9B6}90.48$& $3.33$& $90.19$& $89.28$& $\cellcolor[HTML]{F1B9B6}0.58$& $\cellcolor[HTML]{F1B9B6}90.96$\\
Blended \cite{chen2017targeted} & $86.70$& $8.23$& $93.03$& $85.96$& $7.81$& $92.87$& $90.85$& $9.99$& $94.22$& $89.36$& $\cellcolor[HTML]{F1B9B6}0.07$& $\cellcolor[HTML]{D7E8F2}98.44$& $\cellcolor[HTML]{D7E8F2}90.92$& $\cellcolor[HTML]{D7E8F2}1.46$& $\cellcolor[HTML]{FEEBB8}98.52$& $90.29$& $\cellcolor[HTML]{FEEBB8}0.14$& $\cellcolor[HTML]{F1B9B6}98.87$\\
Input-Aware \cite{nguyen2020input} & $\cellcolor[HTML]{D7E8F2}90.59$& $3.41$& $95.58$& $87.33$& $\cellcolor[HTML]{D7E8F2}1.19$& $96.03$& $87.99$& $2.57$& $95.67$& $89.74$& $1.51$& $\cellcolor[HTML]{D7E8F2}96.53$& $89.99$& $\cellcolor[HTML]{FEEBB8}0.52$& $\cellcolor[HTML]{FEEBB8}97.03$& $90.55$& $\cellcolor[HTML]{F1B9B6}0.48$& $\cellcolor[HTML]{F1B9B6}97.05$\\
LF \cite{zeng2021rethinking} & $74.47$& $2.17$& $51.43$& $85.64$& $1.56$& $56.14$& $87.64$& $\cellcolor[HTML]{F1B9B6}0.91$& $\cellcolor[HTML]{F1B9B6}56.46$& $\cellcolor[HTML]{D7E8F2}89.52$& $\cellcolor[HTML]{D7E8F2}1.32$& $\cellcolor[HTML]{D7E8F2}56.26$& $\cellcolor[HTML]{FEEBB8}89.66$& $2.01$& $55.91$& $\cellcolor[HTML]{F1B9B6}89.93$& $1.38$& $56.23$\\
SSBA \cite{li2021invisible} & $87.65$& $1.84$& $95.03$& $86.86$& $3.03$& $94.04$& $88.94$& $4.21$& $94.50$& $89.13$& $\cellcolor[HTML]{FEEBB8}0.68$& $\cellcolor[HTML]{D7E8F2}96.36$& $\cellcolor[HTML]{F1B9B6}90.28$& $\cellcolor[HTML]{F1B9B6}0.50$& $\cellcolor[HTML]{F1B9B6}97.02$& $89.38$& $1.32$& $96.16$\\
Trojan \cite{Trojannn} & $86.54$& $5.13$& $94.92$& $86.33$& $0.19$& $97.29$& $89.79$& $0.86$& $98.68$& $89.68$& $\cellcolor[HTML]{FEEBB8}0.00$& $\cellcolor[HTML]{FEEBB8}99.06$& $\cellcolor[HTML]{FEEBB8}90.07$& $0.96$& $98.77$& $\cellcolor[HTML]{F1B9B6}90.74$& $\cellcolor[HTML]{F1B9B6}0.00$& $\cellcolor[HTML]{F1B9B6}99.58$\\
WaNet \cite{nguyen2021wanet} & $\cellcolor[HTML]{FEEBB8}90.75$& $1.10$& $\cellcolor[HTML]{D7E8F2}97.70$& $88.24$& $1.72$& $97.38$& $88.25$& $1.88$& $97.31$& $89.09$& $\cellcolor[HTML]{FEEBB8}0.46$& $\cellcolor[HTML]{F1B9B6}98.02$& $89.84$& $\cellcolor[HTML]{D7E8F2}0.47$& $\cellcolor[HTML]{FEEBB8}98.01$& $89.63$& $1.74$& $97.38$\\ \midrule
Average & $86.64$& $3.06$& $89.15$& $86.63$& $2.66$& $89.48$& $89.10$& $2.86$& $90.49$& $89.15$& $\cellcolor[HTML]{FEEBB8}1.04$& $\cellcolor[HTML]{D7E8F2}91.19$& $\cellcolor[HTML]{F1B9B6}90.14$& $\cellcolor[HTML]{D7E8F2}1.40$& $\cellcolor[HTML]{FEEBB8}91.43$& $\cellcolor[HTML]{FEEBB8}89.91$& $\cellcolor[HTML]{F1B9B6}0.78$& $\cellcolor[HTML]{F1B9B6}91.58$\\
\bottomrule
\end{tabular}}
\end{table*}

\section{Defense results in comparison to SOTA defenses on more datasets and networks\label{app_C}}
To demonstrate the effectiveness of our defense methods across different datasets and networks, we conduct additional experiments on the Tiny ImageNet dataset (Tab. \ref{table3}) and on the VGG19-BN network with the CIFAR-10 dataset (Tab. \ref{table4}), comparing their performance against SOTA defenses. When evaluated on Tiny ImageNet (Tab. \ref{table3}), our methods consistently achieve remarkably low Attack Success Rates (ASRs) while maintaining competitive clean accuracy (ACC). For instance, under severe poisoning conditions (10\%), e-CNPD and a-CNPD effectively suppress ASR to near-zero for attacks like BadNets~\cite{gu2019badnets}, Trojan~\cite{Trojannn}, and SSBA~\cite{li2021invisible}. Meanwhile, our methods preserve accuracy at levels comparable to or even surpassing SOTA approaches, demonstrating that they can simultaneously ensure robustness and maintain the model’s utility. Similar trends are observed on VGG19-BN network (Tab. \ref{table4}), where e-CNPD and a-CNPD exhibit consistently low ASRs, indicating that our approaches generalize well across datasets and network architectures. Notably, the DER remains favorable, reflecting a balanced trade-off between backdoor mitigation and classification performance.

\vspace{0.5em}
\noindent \textbf{Effectiveness against state-of-the-art defenses.} Our proposed defenses outperform or match the leading defenses across multiple challenging backdoor attacks. Compared to prominent existing methods such as FP~\cite{liu2018fine}, NAD~\cite{li2021neural}, NC~\cite{wang2019neural}, ANP~\cite{wu2021adversarial}, and i-BAU~\cite{zeng2022adversarial}, our e-CNPD and a-CNPD consistently yield lower ASRs. More importantly, they do so without sacrificing clean accuracy significantly, which underscores their practical relevance. The comparison with defenses like FT-SAM~\cite{zhu2023enhancing}, FST~\cite{min2024towards}, and SAU~\cite{wei2023shared} further consolidates this point. While some existing methods may reduce ASR, they often incur substantial accuracy drops. In contrast, our methods navigate the delicate balance between robustness and accuracy more effectively. Over multiple attack strategies, including newer and more adaptive attacks (e.g., Adap-Blend~\cite{qi2023revisiting}), our defenses maintain superiority or close parity, indicating their resilience against evolving backdoor strategies.

\vspace{0.5em}
\noindent \textbf{Comparison among our four defenses.} Among the proposed approaches, e-CNPD and a-CNPD consistently emerge as the top-performing variants in terms of minimizing ASR while preserving accuracy across both Tiny ImageNet and CIFAR-10. NPD and r-CNPD, though effective, exhibit slightly higher ASRs in certain scenarios compared to e-CNPD and a-CNPD. Nevertheless, NPD and r-CNPD still surpass most baseline and SOTA defenses. E-CNPD generally provides a robust and balanced solution, maintaining low ASR with minimal impact on ACC. A-CNPD further refines this balance, often achieving the lowest ASRs and DER values observed, solidifying its role as a highly competitive defense. Thus, among the three, a-CNPD stands out as the most robust, though e-CNPD remains a strong contender, offering a complementary performance profile that may be preferable in certain application constraints.

In summary, the above analysis shows that our methods consistently achieve low attack success rates while maintaining competitive accuracy. Compared to state-of-the-art defenses, our methods outperform them, demonstrating robustness and utility across various attacks.

\end{document}